\documentclass{ieeeaccess}
\usepackage{cite}
\usepackage{amsmath,amssymb,amsfonts}
\usepackage{graphicx}
\usepackage{soul}
\usepackage{textcomp}
\usepackage{subcaption}
\usepackage{caption}
\usepackage{subcaption}
\usepackage{multirow}
\usepackage{multicol}
\usepackage{adjustbox}
\usepackage{booktabs}
\usepackage{float}
\usepackage{url}
\usepackage{array}
\usepackage{algorithm}
\usepackage{algpseudocode}
\usepackage{hyperref}
\usepackage{stfloats}
\usepackage{listings}
\definecolor{lightgoldenrodyellow}{RGB}{250, 250, 210}
\sethlcolor{lightgoldenrodyellow}
\def\BibTeX{{\rm B\kern-.05em{\sc i\kern-.025em b}\kern-.08em
    T\kern-.1667em\lower.7ex\hbox{E}\kern-.125emX}}
\begin{document}
\history{Date of publication xxxx 00, 0000, date of current version xxxx 00, 0000.}
\doi{10.1109/ACCESS.2024.DOI}

\title{
BlockMEDC: Blockchain Smart Contracts for Securing Moroccan Higher Education Digital Certificates
}

\author{\uppercase{Mohamed Fartitchou}\authorrefmark{1}, \uppercase{Ismail Lamaakal}\authorrefmark{1}, (Member, IEEE), 
\uppercase{Khalid El Makkaoui}\authorrefmark{1}, (Senior Member, IEEE), \uppercase{Zakaria El Allali}\authorrefmark{1}, \uppercase{Yassine Maleh}\authorrefmark{2}, (Senior Member, IEEE)}
\address[1]{Multidisciplinary Faculty of Nador, Mohammed Premier University, Oujda 60000, Morocco}
\address[2]{Laboratory LaSTI, ENSAK, Sultan Moulay Slimane University, Khouribga 54000, Morocco}


\markboth
{M. Fartitchou \headeretal: BlockMEDC: Blockchain Smart Contracts for Securing Moroccan Higher  Education Digital Certificates}
{M. Fartitchou \headeretal: BlockMEDC: Blockchain Smart Contracts for Securing Moroccan Higher Education Digital Certificates}



\corresp{Corresponding author: Yassine Maleh (e-mail: yassine.maleh@ieee.org).}

\begin{abstract}
Morocco’s Vision 2030, known as Maroc Digital 2030, aims to position the country as a regional leader in digital technology by boosting digital infrastructure, fostering innovation, and advancing digital skills. Complementing this initiative, the Pacte ESRI 2030 strategy, launched in 2023, seeks to transform the higher education, research, and innovation sectors by integrating state-of-the-art digital technologies.
In alignment with these national strategies, this paper introduces BlockMEDC, a blockchain-based system for securing and managing Moroccan educational digital certificates. Leveraging Ethereum smart contracts and the InterPlanetary File System, BlockMEDC automates the issuance, management, and verification of academic credentials across Moroccan universities. The proposed system addresses key issues such as document authenticity, manual verification, and lack of interoperability, delivering a secure, transparent, and cost-effective solution that aligns with Morocco's digital transformation goals for the education sector.
\end{abstract}


\titlepgskip=-15pt
\clearpage 
\maketitle
\section{Introduction} 
\label{sec1}

Morocco's Digital 2030 (MD2030) Vision aims to harness global digital advancements to transform the kingdom into a prominent producer of digital solutions and a regional leader in technology and innovation. Officially launched on September 25, 2024 \cite{r116}, the National Strategy Digital Morocco 2030 outlines a strategic initiative focused on advancing digital infrastructure by digitizing public services, supporting and enhancing start-ups, developing digital skills, providing high-quality internet coverage across Morocco, emphasizing computing and artificial intelligence, and promoting international collaboration \cite{b01}.

In alignment with MD2030 Vision, Morocco's Pacte ESRI 2030, launched in 2023, aims to modernize higher education, research, and innovation sectors by integrating advanced digital technologies and innovative practices to better align with global advancements and address contemporary challenges and pandemics (e.g., COVID-19) \cite{b02, b03}.

Recently, e-learning platforms have become pivotal in Moroccan higher education, complementing face-to-face teaching with advanced digital tools, e.g., Moodle and Rosetta Stone \cite{b04, b04b, b05}.  Moodle provides tools for course creation, assignment management, and communication between students and instructors. Rosetta Stone is used primarily for foreign language learning, which helps learners develop their skills through interactive lessons and exercises.

 As digital technologies continue to evolve, several Moroccan universities are transitioning to digital certifications, including Mohammed First University of Oujda, Mohammed V University of Rabat, Sultan Moulay Slimane University of Beni Mellal, and Ibn Zohr University of Agadir. These certifications ensure document validity through digital signatures and can be easily accessed, shared, and verified online~\cite{b06}.

\subsection{Overview of Digital Certificates}
 Moroccan universities' adoption of digital certificates (e.g., registration certificates, official academic transcripts, and diplomas) aims to modernize and streamline the process of obtaining and verifying academic certificates. In what follows, we overview the current state of Moroccan higher education's digital certificate system. Firstly, we outline the procedures for requesting and issuing digital certificates. Next, we highlight the digital signature process. Finally, we discuss the digital certificate system's benefits and issues.
\subsubsection{Process of requesting and issuing}
\begin{figure}[h!]
    \centering
    \includegraphics[scale=0.3]{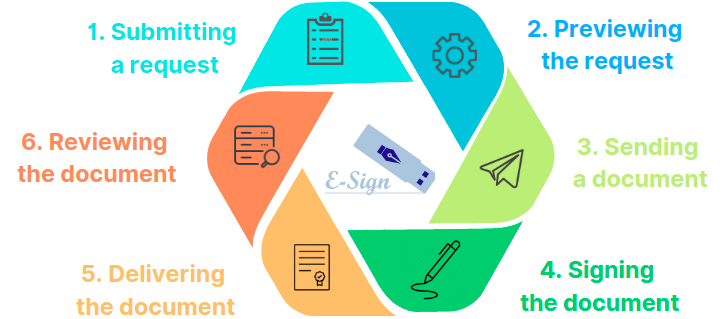}
    \caption{The stages of issuing digital certificates.}
    \label{fig1}
\end{figure}

 Figure~\ref{fig1} illustrates the stages involved in processing digital certificate requests in the current Moroccan system, from submitting a request to reviewing the document.

\begin{figure}[h!]
    \centering
    \includegraphics[scale=0.5]{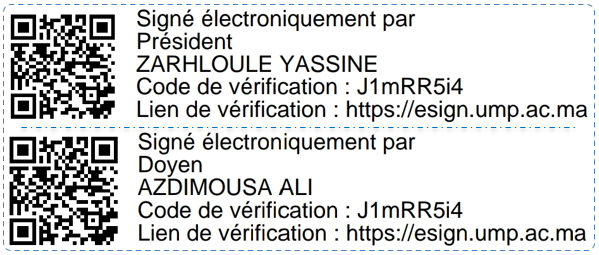}
    \caption{A diploma electronically signed by the president and the dean.}
    \label{fig2}
\end{figure}

Where, 
\begin{itemize}
    
    \item [\textbf{1.}]  \textbf{Submitting a request}:  A student can use their platform account to request a document, except for the diploma, which is issued automatically upon validation for all semesters.
    \item [\textbf{2.}]  {\textbf{Previewing the request}: The administrator of digital certificates inspects the received request.}
    \item [\textbf{3.}]  {\textbf{Sending a document}: The administrator sends the required document for signing.}
    \item [\textbf{4.}]  {\textbf{Signing the document}: The authorized signatory (e.g., the university president and/or dean) receives the document electronically and applies their digital signature. This electronic signature is typically appended to the end of the document, as illustrated in Figure~\ref{fig2}.}
    \item [\textbf{5.}]  {\textbf{Delivering the document}: The student receives the stamped document through their account on the platform.}
    \item [\textbf{6.}]  \textbf{Reviewing the document}: A central server stores all signed documents. Anyone can verify the authenticity of the document issuance using a special link and verification code (e.g., \texttt{link}: \url{https://esign.ump.ac.ma} and \texttt{code}: J1mRR5i4, see Figure~\ref{fig2}).
\end{itemize}
\subsubsection{Digital signature}
Digital signatures are a crucial element in ensuring the authenticity and integrity of digital certificates. They rely on cryptographic algorithms, specifically using a pair of keys:
\begin{itemize}
    \item \textbf{Private Key:} Kept confidential by the signer (e.g., university president).
    \item \textbf{Public Key:} Made available for verification purposes.
\end{itemize}

Morocco's Law n° 43-20 (Dahir n° 1-20-100 of December 31, 2020) mandates the use of confidence services, including digital signatures, to secure digital transactions \cite{b07}. Barid eSign \cite{b08}, the sole state-approved certification authority (CA), provides essential services such as digital signatures, strong authentication, timestamps, and SSL certificates to ensure the integrity and trustworthiness of digital communications in Morocco.

Each signer receives a SecurID key from Barid eSign, which contains certified keys. The signing process is as follows:
\begin{enumerate}
    \item  Authorized signers (e.g., presidents, deans, and directors) use their RSA-2048 \cite{b09} private keys, which are stored securely in a SecurID key and accessed with a PIN, to create a digital signature.
    \item The signature incorporates either the SHA-1 or SHA-256 hash algorithm \cite{b010} to ensure data integrity.
    \item  This digital signature is embedded into the digital certificate.
\end{enumerate}

The missing link in current Moroccan digital certificates is the secure verification process, which depends on the following:
\begin{enumerate}
    \item Entering the certificate's reference number on the institution's e-sign platform to initiate verification.
    \item Contacting Barid eSign to confirm the signer's identity details (name, public key, hash function).
    \item Utilizing the provided public key and hash function to independently verify the integrity of the signed certificate.
\end{enumerate}

\subsubsection{Benefits}
 Digital certificates offer a range of benefits to both individuals and organizations in the Moroccan higher education sector, including:

\begin{itemize}
\item  {\textbf{Accelerated processes}: By automating many administrative tasks, the digital certificate system enables educational institutions to issue certificates more quickly than traditional paper-based systems. These tasks include submitting online requests, verifying student requests, digitally signing certificates, and delivering the certificates electronically.}
\item  {\textbf{Reduced costs}: Using digital certificates can significantly reduce costs for both students and educational institutions in terms of time and money, as they eliminate the need for printing (except diplomas), in-person delivery, and physical storage of paper certificates. Furthermore, students will no longer need to cast certified copies of the original certificates because the current system has set up an online central server to verify certificate authenticity.}
\item  {\textbf{Environmentally friendly}: Digital certificates help promote sustainable practices by reducing the use of paper, ink, energy, and transportation.}
\end{itemize}
\subsubsection{Issues}
 {Despite the many advantages that digital certificates offer over their paper-based counterparts, there are still a number of security and efficiency-related issues that require attention, including:}

\begin{itemize}
 \item  { \textbf{Security threats}: Both digital and traditional certificate systems are susceptible to forgery because the administrators have access to all student grades and can potentially manipulate them. Once the grades are submitted to the administrator, it becomes challenging for the professors to monitor the integrity of the grades. On the other hand, the central server, which stores all signed documents, is vulnerable to various attacks, including denial of service; the most dangerous of them are forging diplomas and adding another that does not exist.}
\item  \textbf{Hand-to-hand distribution:}  It requires students to physically visit the institution to obtain a printed and stamped copy of their digital diplomas. This process is costly for both students and the diploma desk, as it consumes a significant amount of time and resources, resulting in a lengthy and inefficient distribution process.
 \item  {\textbf{Manual verification}: This is an issue with the current digital certificate system. Verifying the authenticity of a digital certificate requires downloading it from a central server using a special link and verification code. This manual process can be time-consuming and prone to errors.}
\item  {\textbf{Semi-automated}: The current digital certificate system does not fully automate the issuance process. Students must request the required documents (except diplomas), and administrators must review any received requests.}
\item  {\textbf{Not instantaneous}: The issuance of official academic transcripts and diplomas is not an instantaneous process; it can take more than a month after the announcement of regular session results.}
\item  {\textbf{Non-uniform profile}: The current Moroccan digital certificate system lacks consistency in creating academic accounts for students, as each university generates accounts independently. This can lead to students having multiple accounts if they receive certificates from different universities, resulting in a fragmented and potentially confusing digital profile.}

\end{itemize}
\subsection{Research Motivation} 
 {The urgent need to improve the security, efficiency, and interoperability of the current digital certificate system in Moroccan higher education, while aligning with MD2030 Vision and Pacte ESRI 2030 objectives, motivates this research. The integration of blockchain smart contracts \cite{b011, b012, b012a, b012b, b012c, b012d, b013, b014, b015, b016, b017, b017b} offers a decentralized, secure, and tamper-proof framework that has the potential to revolutionize the educational digital certificate system, enhancing several key aspects:}

\begin{itemize}
    \item  {\textbf{Security and Integrity:} BC offers a robust defense against forgery and unauthorized alterations, ensuring that digital certificates remain secure and trustworthy.}
    \item  {\textbf{Transparency and Traceability:} The transparent nature of BC allows for real-time verification of certificates, creating a clear audit trail and fostering trust among stakeholders.}
    \item  {\textbf{Efficiency and Automation:} Smart contracts (SCs) automate the issuance and verification processes, reducing administrative burdens and expediting operations.}    
    \item  {\textbf{Interoperability and Standardization:} A BC-based platform enables standardized digital certificates, promoting consistency and interoperability across different educational institutions.}
    \item  {\textbf{Cost Reduction and Environmental Sustainability:} By automating processes and eliminating paper-based documents, BC technology reduces costs and supports environmentally sustainable practices for educational institutions, students, and professors.}
\end{itemize}

\subsection{Main Contributions}
 This research, for the first time, proposes a system based on Ethereum SCs designed for securing and managing digital education certificates within the Moroccan context. This novel system, referred to as BlockMEDC, presents several original contributions, outlined below:
\begin{itemize}
    \item  { \textbf{Remote Deliberation:} BlockMEDC facilitates the remote deliberation of student grades across both regular and catch-up exam sessions, leveraging a decentralized Ethereum network. By transcending geographical constraints, this approach enhances the efficiency and accessibility of the certification process.}
    
    \item   {\textbf{Automation and Real-Time Issuance:}  BlockMEDC uses Ethereum SCs to automate digital certificate issuance and management. Once the conditions for their issuance are met, all documents—such as registration certificates, official academic transcripts, and diplomas—are generated and issued automatically and in real time. } 
     \item  {\textbf{Issuing Digital Certificates for Professors}: It is a feature of BlockMEDC that extends the platform's functionality to include the secure issuance, management, and verification of digital certificates for professors, such as university qualifications and teaching credentials. This approach simplifies the verification process for employment, promotions, and academic collaborations, empowering professors to manage and share their credentials easily.}   
    \item   {\textbf{Ensuring Integrity:} BlockMEDC guarantees the integrity of students' grades, as well as both students' and professors' certificates, making them resistant to tampering and unauthorized alterations. Additionally, the system enables professors to monitor the integrity of students' grades both before and after the deliberation process.}
    \item   {\textbf{Standardization Across Institutions:} The BlockMEDC system introduces standardization across Moroccan educational institutions by providing each student/professor with a single, unified academic account. This approach eliminates account fragmentation, simplifies record management, and enhances the integration and usability of students' and professors' academic information across different universities.}
    \item   \textbf{Interoperability:} BlockMEDC can efficiently interact with different platforms, facilitating and automating the secure verification of digital certificates across diverse institutions. For example, to participate in the master's or PhD entrance exam, a student can simply provide their BlockMEDC\_ID, allowing stakeholders to securely access and verify their academic records.
    \item   {\textbf{Economic and Environmental Benefits:} The proposed system promotes cost reduction and environmental sustainability by automating certificate issuance and verification processes, eliminating physical documentation, and reducing energy consumption.}    
\end{itemize}

\subsection{Organization of the Paper}
This paper is structured as follows: Section~\ref{sec2} presents a review of existing blockchain-based digital certificate systems, discussing their key features and identifying their strengths and limitations. Section~\ref{sec3} introduces the foundational concepts and technologies underlying BlockMEDC, including Ethereum, smart contracts, and IPFS. Section~\ref{sec4} outlines the design and architecture of the BlockMEDC system, explaining its components, workflow, and the role of smart contracts. Section~\ref{sec5} presents the implementation setup and results, detailing the deployment environment, test network, and evaluation of transaction costs. Finally, Section~\ref{sec6} concludes the paper, summarizing the key findings and highlighting future research directions to enhance system scalability, optimize transaction costs, and improve privacy through the use of zero-knowledge proofs.

\section{Related Work} 
\label{sec2}
Recent years have seen the proposal and development of several BC-based certificate platforms and systems, such as those at the University of Nicosia \cite{b018}, MIT \cite{b019, b020}, and Open University \cite{b021}. These platforms and systems leverage the security and transparency of the BC technology to address the challenges faced by traditional and digital certificate systems, such as fraud, forgery, and inefficiency. This section reviews the existing literature and related work on BC-based digital certificate systems for education, highlighting their strengths and weaknesses.

In \cite{b024}, the authors proposed a novel BC-based education records verification solution aimed at enhancing the efficiency and security of managing educational credentials. By utilizing a decentralized architecture, the system allows individuals to control their educational data while facilitating seamless sharing among authorized institutions, thereby eliminating the need for intermediaries and reducing verification costs. The proposed solution incorporates a consensus algorithm to ensure the validity of transactions, enabling secure access to records through individual private keys.

The authors of \cite{b023} proposed a BC-based system using Ethereum and Solidity to securely manage academic records and automate the issuance of degree certificates in Brazil. The system employs three SCs—authority, curriculum, and diploma—to prevent fraud and inefficiencies, ensuring that only authorized institutions can validate and issue certificates. The study highlights the potential of BC to improve the security and transparency of academic credential management, address privacy concerns and costs, and offer a model for global application.

In \cite{b022}, the authors proposed a decentralized system for managing educational certificates using the Ethereum BC and smart contracts (SCs). The system securely manages certificate issuance, verification, and revocation through SCs, ensuring transparency, tamper resistance, and decentralization. It employs multiple specialized contracts, each responsible for a different role in the certificate lifecycle. The integration of cryptographic algorithms and SC logic enhances security, eliminating reliance on third-party systems and preventing unauthorized access. The system's design also allows for compatibility with existing protocols, making it flexible for broader applications.

The authors of \cite{b025} proposed a BC-based framework for the verification of educational certificates to address significant security gaps in existing solutions, specifically focusing on themes such as authentication, authorization, confidentiality, privacy, and ownership. Through a comprehensive review of current certification solutions, they identified that many inadequately addressed these critical aspects, leaving certificates vulnerable to fraud. By leveraging the Hyperledger Fabric framework, the proposed system aims to enhance the verification process.

The authors of \cite{b028} proposed a novel architecture called HEDU-ledger, utilizing Hyperledger Fabric technology to create a private permissioned BC network aimed at enhancing the security and traceability of degree attestation and verification processes in higher education. This architecture addresses critical issues such as data tampering, forgery, and the inefficiencies of traditional centralized systems by providing a decentralized, immutable ledger for storing academic credentials.

The authors of \cite{b027} proposed an implementation model for an academic degree certification system utilizing BC technology, specifically through the start-up BeCertify. This model aims to address the challenges of data management and accessibility in higher education by providing a decentralized, secure, and transparent method for issuing and verifying academic credentials. The paper outlines the various stages of development for the platform, including system architecture and REST API operations, while also discussing the multi-signature authorization capabilities that enhance security and user control.

In \cite{b026}, the authors proposed a BC-based certification system for academic degrees through the start-up BeCertify, aiming to enhance the transparency, security, and accessibility of educational records. They outlined the various stages of development necessary for the platform, including system architecture and REST API operations, while addressing challenges encountered during implementation. The study highlighted the importance of multi-signature authorization and the decentralized nature of blockchain, which empowers students and educational institutions to manage academic records more effectively.

In \cite{b029}, the authors proposed a comprehensive BC-based solution named Cerberus for efficient verification of academic credentials, addressing the pervasive issue of credential fraud in higher education.
By integrating seamlessly with existing credential management ecosystems, the system utilizes a permissioned BC maintained by accreditation bodies and universities, allowing for real-time verification through QR codes linked to graduates' records. 

The authors of \cite{b030} proposed a BC-based framework for the secure and efficient verification of educational credentials, utilizing Ethereum smart contracts and the IPFS for decentralized data storage. This model aims to address prevalent issues in the education sector, such as the risk of academic fraud and the inefficiencies associated with traditional verification processes. By implementing a method that randomizes credential attribute values and constructs a verification tree, the system enhances privacy and allows for selective disclosure of information.

The authors of \cite{b031} proposed a pilot project aimed at verifying student ID information and transcripts of records between KU Leuven and Università di Bologna, utilizing the European Blockchain Services Infrastructure (EBSI) to enhance the digital exchange of educational credentials. This initiative is part of the Una Europa network, which seeks to streamline and secure the verification process for exchange students by issuing verifiable digital identities and transcripts. Through action research, the study assessed the institutional and technical conditions necessary for scaling the pilot, revealing that while the EBSI framework met initial design principles, additional capabilities are required for full production implementation.

Table \ref{tab1} provides a comparative analysis of existing BC-based educational credential systems, highlighting their proposed solutions, key features, strengths, weaknesses, and the integration of smart contracts (SCs).

\begin{table*}[htbp]
\centering
\caption{Comparison of Related Work and Proposed System}
\begin{adjustbox}{width=\linewidth}
\begin{tabular}{|p{1.5cm}|c|p{4cm}|p{4cm}|p{4cm}|p{4cm}|c|}
\hline
\textbf{Ref.} & \textbf{Year} & \textbf{Proposed Solution} & \textbf{Key Features} & \textbf{Strengths} & \textbf{Weaknesses} & \textbf{SCs} \\
\hline
\multirow{3}{*}{\cite{b024}} & \multirow{3}{*}{2018} & Decentralized environment for individuals to manage their education records & Data privacy, data security, decentralized data storage, customized access & Secure, reduces verification costs, supports lifelong education & Potential data management issues, requires user familiarity with blockchain technology & \multirow{3}{*}{\includegraphics[scale=0.2]{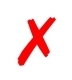}} \\
\hline
\multirow{3}{*}{\cite{b023}} & \multirow{3}{*}{2019 }& Blockchain and smart contracts for higher education registry in Brazil & Securely manage academic records and automate degree issuance & Prevents fraud and inefficiencies, ensures authorized institutions validate certificates & Requires significant technical expertise for implementation, privacy concerns & \multirow{3}{*}{\includegraphics[scale=0.2]{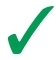}} \\
\hline
\multirow{3}{*}{\cite{b022}} & \multirow{3}{*}{2020 }& Blockchain for Education: Lifelong Learning Passport & Decentralized system for managing educational credentials & Secure, transparent, tamper-resistant, supports lifelong learning & May require changes to existing educational infrastructure, potential scalability issues & \multirow{3}{*}{\includegraphics[scale=0.2]{yes.jpg}}\\
\hline
\multirow{3}{*}{\cite{b025}} & \multirow{3}{*}{2020} & Blockchain-based framework using Hyperledger Fabric & Authentication, authorization, confidentiality, privacy, ownership & Enhances verification process, addresses security gaps & May be complex to implement, relies on Hyperledger Fabric, potential scalability issues & \multirow{3}{*}{\includegraphics[scale=0.2]{no.jpg}} \\
\hline
\multirow{3}{*}{\cite{b028}} & \multirow{3}{*}{2021} & Private permissioned blockchain for degree attestation and verification & Security, traceability, decentralized ledger & Secure, addresses data tampering and forgery issues & Requires permissioned blockchain setup, potential implementation complexity & \multirow{3}{*}{\includegraphics[scale=0.2]{no.jpg}} \\
\hline
\multirow{3}{*}{\cite{b027}} & \multirow{3}{*}{2022}& Blockchain implementation model for academic degree certification & Multi-signature authorization, decentralized method, REST API operations & Secure, transparent, enhances data management & Implementation challenges, requires significant technical expertise & \multirow{3}{*}{\includegraphics[scale=0.2]{yes.jpg}} \\
\hline
\multirow{3}{*}{\cite{b026}} & \multirow{3}{*}{2023} & Blockchain-based solution for academic credential verification & Real-time verification, QR code integration, permissioned blockchain & Reduces credential fraud, integrates with existing ecosystems & May require significant changes to existing systems, potential cost implications & \multirow{3}{*}{\includegraphics[scale=0.2]{no.jpg}} \\
\hline
\multirow{3}{*}{\cite{b029}} & \multirow{3}{*}{2023} & Blockchain-based solution for academic credential verification & Real-time verification, QR code integration, permissioned blockchain & Reduces credential fraud, integrates with existing ecosystems & May require significant changes to existing systems, potential cost implications & \multirow{3}{*}{\includegraphics[scale=0.2]{yes.jpg}}\\
\hline
\multirow{3}{*}{\cite{b030}} & \multirow{3}{*}{2023} & Blockchain-based framework using Ethereum and IPFS for secure credential verification & Randomized credential attribute values, verification tree, selective disclosure & Enhances privacy, supports decentralized storage, secure & Implementation complexity, reliance on Ethereum and IPFS & \multirow{3}{*}{\includegraphics[scale=0.2]{yes.jpg}} \\
\hline
\multirow{3}{*}{\cite{b031}} & \multirow{3}{*}{2023} & Pilot project for verifying student ID information and transcripts using EBSI & Digital identities, verifiable credentials, institutional and technical assessments & Supports digital exchange, enhances verification process & Requires further development for full production, potential scalability issues & \multirow{3}{*}{\includegraphics[scale=0.2]{no.jpg}} \\
\hline
\textbf{Our Proposed System} & \multirow{3}{*}{2024} & Decentralized digital certificate system using blockchain smart contracts & Automatic generation, decentralized storage, Ethereum network, IPFS integration & Secure, transparent, tamper-proof, real-time updates, supports multiple actors & Initial setup and deployment costs, potential scalability issues, requires user familiarity with BC technology & \multirow{3}{*}{\includegraphics[scale=0.2]{yes.jpg}} \\
\hline
\end{tabular}
\end{adjustbox}
\label{tab1}
\end{table*}

\section{Proposed BlockMEDC System} 
\label{sec3}
This section details the design and architecture of BlockMEDC, a decentralized system based on Ethereum BC and IPFS technologies.
The following subsections provide a comprehensive overview of the various components and modules that constitute BlockMEDC, highlighting how each element contributes to the system's overall functionality and security.

\subsection{System Overview}
\begin{figure}[htbp]
    \centering
    \includegraphics[scale=0.55]{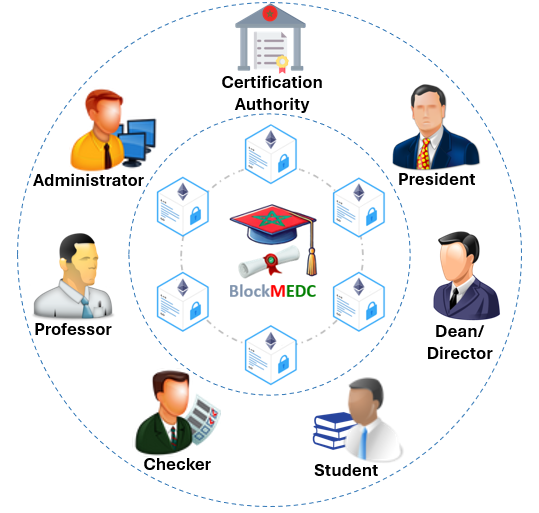}
    \caption{The key actors in our BlockMEDC system.}
    \label{fig3}
\end{figure}

We designed the BlockMEDC system to enhance the security, efficiency, and accessibility of digital certificate issuance and verification within Morocco's higher education institutions. Using Ethereum BC technology, BlockMEDC automates the management and real-time issuance of academic digital documents, such as registration certificates, transcripts, and diplomas.
The system allows for remote deliberation of student grades, thereby addressing the impact of geographical constraints on the certification process. BlockMEDC ensures the integrity of academic records by making them resistant to tampering and unauthorized alterations. It also introduces a standardized approach across educational institutions, providing each student with a unified academic account to streamline record management. Additionally, the proposed system offers interoperability with other platforms, enabling secure verification of certificates across diverse institutions.
In the BlockMEDC system, six key actors play essential roles in ensuring the secure and efficient management of digital certificates, including the certification authority (e.g., Barid eSign), president, dean/director, administrator, professor, and student, as well as an external checker responsible for verifying that the digital certificates provided to other parties are accurate, correct, and tamper-proof, as illustrated in Figure~\ref{fig3}.

Table~\ref{tab:roles_responsibilities} outlines the roles and responsibilities of each of BlockMEDC's seven actors.

\renewcommand{\arraystretch}{1} 
\begin{table*}[!h]
    \centering
    \caption{Roles and Responsibilities of BlockMEDC's Key Actors}
    \begin{tabular}{|p{2.5cm} | p{12cm}|}
        \hline
        \textbf{Role} & \textbf{Responsibilities} \\ \hline
        \multirow{4}{*}{Certification Authority} & 
        \begin{itemize}
            \item Directly access the Ethereum network for secure interactions.
            \item Authorize and add university presidents to the BlockMEDC system.
             \item Maintain and manage access control to the system.
        \end{itemize}
        \\ \hline
          \multirow{5}{*}{President} & 
          \begin{itemize}
              \item Authorize and add deans of faculties, directors of schools, and their representatives to the system.
              \item Authorize and add university administrators to the BlockMEDC network.
              \item Approve and validate key academic digital certifications, such as diplomas, before they are added to the network.
          \end{itemize}
          \\ \hline
        \multirow{5}{*}{Dean/Director} & 
        \begin{itemize}
            \item Authorize and add institute administrators to the BlockMEDC system.
            \item Approve and validate all academic digital certifications for their institute.
            \item Manage validated certifications by either adding them directly to the BlockMEDC system or sending them to the university president for further approval.
        \end{itemize}
        \\ \hline
       \multirow{5}{*}{Administrator} & 
       \begin{itemize}
            \item  Manage the system's operations to ensure all processes run smoothly.
            \item Maintain the normal functioning of the BlockMEDC platform.
            \item Create and manage professor, student, and other administrator accounts.
        \end{itemize}
        
        \\ \hline
         \multirow{8}{*}{Professor} & 
        \begin{itemize}
            \item Securely enter and verify student grades on the decentralized BlockMEDC platform.
            \item Participate in remote deliberation sessions to discuss and finalize student grades with other faculty members.
            \item Serve as a department head or master’s program head with the authority to approve and validate certain academic certifications.
            \item Manage, download, and share their academic digital certificates, such as university qualification certificates.
        \end{itemize}
        
        \\ \hline
        \multirow{6}{*}{Student} & 
        \begin{itemize}
            \item Download, manage, and securely share their academic records with relevant parties, such as institutions or employers, on the BlockMEDC platform.
            \item Submit a claim to request a review or correction if they believe there has been an error, such as an incorrect grade.
            \item Request the addition of extra modules as needed.
        \end{itemize}
        
        \\ \hline
        \multirow{6}{*}{Checker} & 
       \begin{itemize}
           
            \item For interoperable platforms, ensure the verification process is implicit and automatic.
            \item For non-interoperable platforms, manually access the BlockMEDC platform to upload and verify certificates.
        \end{itemize}
        
        \\ \hline
    \end{tabular}
    \label{tab:roles_responsibilities}
\end{table*}

\begin{figure*}[htbp]
\begin{center}
\includegraphics[scale=0.4]{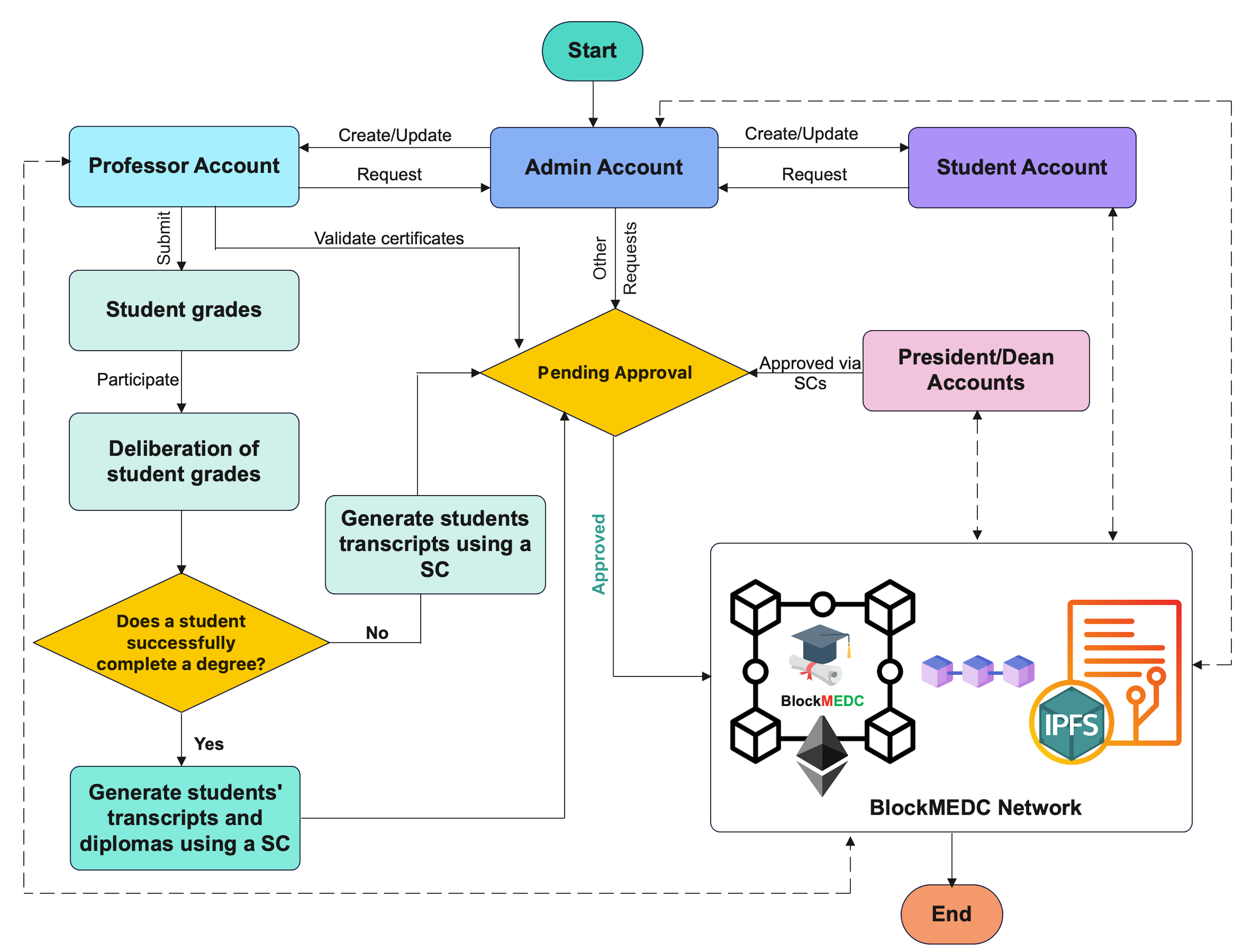}
\caption{BlockMEDC system flowchart.}
\label{flowchart}
\end{center}
\end{figure*}

Once the BlockMEDC accounts for presidents, deans/directors, and administrators are created, they can oversee the process of issuing digital certificates. 
The flowchart (Figure~\ref{flowchart}) illustrates the process of managing professor and student accounts, including their creation and updates by administrators at each institute. It also details the step-by-step process of generating, approving, validating, and issuing academic digital certificates within the BlockMEDC system; the following steps are involved:

\begin{enumerate} \item The administrator generates all necessary digital certificates, except for students' transcripts and diplomas, after creating or updating student and professor accounts using smart contracts. All generated certificates are pending approval. \item Students have the option to request additional course units or to change their academic path/major. \item Professors who serve as department heads or master’s program heads are responsible for validating specific academic certifications; the validation can be done using a smart contract. \item Professors have the option to request extra certifications. \item Each professor submits the grades of their students. The system allows professors to ensure that these grades remain immutable and protected against any unauthorized alterations. \item Professors participate in remote deliberation sessions for each semester, including both regular and catch-up sessions. To manage student grade deliberation, the BlockMEDC system uses a smart contract that identifies instances where grades necessitate redemption and alerts the concerned teachers for further action. During a normal session, the smart contract detects students who require additional marks to obtain honors. In contrast, during the catch-up session, the smart contract detects all cases where students require additional marks to validate the semester or obtain honors. \item If a student successfully completes a degree, a smart contract generates the associated transcripts and diplomas. Otherwise, the smart contract generates only the associated transcripts. All generated certificates are pending for approval. \item In parallel, the administrator submits the list of students referred to the disciplinary council, as well as the sanctions taken against them. The penalties can include removing some student grades and depriving students of the opportunity to pass certain sessions. \item Students can submit a request to the professors to review and correct their exam papers if they believe that an error has occurred. If the professors detect errors, they resubmit the corrected student grades for remote deliberation. \item Before sending the digital certificates to the BlockMEDC network, both the institution head and the university president digitally sign the certificates using smart contracts, with each signing according to their respective responsibilities. \item The Ethereum network validates all approved transactions to issue the certificates, with the hash of the issued certificates being added to the Ethereum blockchain, while the actual files are stored on the IPFS network. \end{enumerate}

\section{Ethereum-Based Smart Contracts Architecture}
\label{sec4}

This section details the prototype implementation of our proposed solution, which leverages the Ethereum blockchain platform and Solidity to create the smart contracts (SCs) introduced in the preceding section.

\subsection{Ethereum Network for Blockchain Integration}
We discuss the rationale for selecting the Ethereum network as the underlying blockchain technology for our digital certification system and present a comparative analysis of Ethereum against other prominent blockchain platforms to highlight its suitability.

\textbf{Ethereum} is a decentralized, open-source blockchain platform that empowers developers to create and deploy decentralized applications (dApps) and smart contracts. \cite{r111,r112} Introduced in 2015 by Vitalik Buterin, Ethereum introduced a programmable blockchain model that goes beyond simple transactions. With its built-in programming language, Solidity, Ethereum enables the development of self-executing smart contracts, which automate and enforce agreements without the need for intermediaries.

Operating on a peer-to-peer network, Ethereum guarantees data transparency and integrity across its decentralized ledger. Recently, Ethereum transitioned from the energy-intensive Proof of Work (PoW) consensus mechanism to a more eco-friendly and scalable Proof of Stake (PoS) model \cite{r123}. This evolution has further solidified Ethereum's position as a leading platform for a wide range of use cases, including digital identity management, supply chain tracking, and digital certification systems. \\

\textbf{The InterPlanetary File System} is a decentralized storage solution designed to store and share data across a distributed network \cite{r113,r115}. Unlike traditional centralized servers, IPFS uses a peer-to-peer model, where data is broken into smaller chunks, stored on multiple nodes, and linked together through unique cryptographic hashes called Content Identifiers (CIDs) \cite{r114}. This decentralized architecture ensures data integrity, reduces reliance on centralized storage providers, and enables seamless access to data even if some nodes are offline.

IPFS is particularly beneficial for blockchain applications like Ethereum, as it allows for off-chain storage of large files and data while only storing critical reference information on the blockchain. This combination helps optimize storage costs, improve performance, and enhance the scalability of blockchain-based systems.

\subsubsection{Why We Chose Ethereum for Our Digital Certification System}

After a thorough evaluation of several blockchain platforms, we selected Ethereum as the foundation for our digital certification system due to the following key reasons:

\begin{itemize}
    \item \textbf{Decentralization and Trust}: Ethereum's decentralized structure ensures that no single entity has control over the entire network, which is critical for maintaining trust and transparency in a digital certification system. This decentralized nature eliminates the risk of tampering and ensures that digital certificates remain authentic and secure.

    \item \textbf{Immutability and Security}: Data recorded on the Ethereum blockchain cannot be modified or deleted, making it an ideal platform for storing digital certificates. This immutability guarantees the long-term integrity and authenticity of the certificates, ensuring they are verifiable and resistant to forgery or unauthorized alterations.

    \item \textbf{Smart Contract Automation}: Ethereum supports smart contracts, which enable us to automate processes like certificate issuance, validation, and revocation. This automation reduces the need for manual intervention, minimizes errors, and ensures that all transactions are executed transparently and according to predefined rules.

    \item \textbf{Scalability and Efficiency}: With Ethereum’s transition to the Proof of Stake (PoS) model, the platform has significantly improved its scalability and energy efficiency. This makes Ethereum capable of handling large-scale applications such as our digital certification system, where fast transaction speeds and reduced operational costs are essential.

    \item \textbf{Integration with Off-Chain Solutions}: Ethereum’s compatibility with off-chain storage solutions like the IPFS allows us to store large datasets off-chain while keeping only critical information on the blockchain. This approach reduces the storage burden and enhances system performance and scalability. IPFS ensures that even the largest files, such as complete academic transcripts, are stored securely and can be accessed efficiently without overwhelming the blockchain.

    \item \textbf{Robust Ecosystem and Community Support}: Ethereum’s extensive ecosystem provides a wide range of development tools, libraries, and resources, which simplifies our development process and ensures that we build a secure and reliable system. The active developer community also means that we have access to ongoing support, updates, and improvements to the platform.

    \item \textbf{Proven Track Record}: As one of the earliest blockchain platforms to implement smart contracts, Ethereum has a proven history of supporting complex applications. Its maturity and stability make it a reliable choice for deploying a digital certification system that meets the security and transparency needs of educational institutions.
\end{itemize}

\subsubsection{Comparative Analysis of Blockchain Platforms}

To substantiate our choice, we conducted a comparative analysis of Ethereum against other blockchain platforms such as Hyperledger Fabric \cite{r117}, Stellar \cite{r118}, and EOS \cite{r119}. The comparison is based on criteria essential for digital certification systems, including consensus mechanisms, decentralization, transaction speed, cost efficiency, data privacy, and more (see Table \ref{t100}).

\renewcommand{\arraystretch}{1.4} 
\begin{table*}[htbp]
\centering
\caption{Comparison of Blockchain Platforms for Digital Certification Systems}
\resizebox{\textwidth}{!}{%
\begin{tabular}{|p{2.5cm}|p{2cm}|p{3cm}|p{2.5cm}|p{2cm}|p{2cm}|p{2cm}|p{2cm}|}
\hline
\textbf{Criteria}                & \textbf{Ethereum}       & \textbf{Hyperledger Fabric} & \textbf{Stellar}       & \textbf{EOS}           & \textbf{Corda \cite{r120}}        & \textbf{Tezos \cite{r121}}        & \textbf{Hyperledger Sawtooth \cite{r122}}  \\ \hline
\textbf{Consensus Mechanism}     & Proof of Stake (PoS)    & Practical Byzantine Fault Tolerance (PBFT) & Stellar Consensus Protocol (SCP) & Delegated Proof of Stake (DPoS) & Raft/IBFT            & Liquid Proof of Stake (LPoS) & Proof of Elapsed Time (PoET)   \\ \hline
\textbf{Smart Contract Support}  & \checkmark              & \checkmark                 & \texttimes            & \checkmark            & \checkmark            & \checkmark            & \checkmark            \\ \hline
\textbf{Decentralization Level}  & \checkmark              & \texttimes                 & \checkmark            & \texttimes            & \texttimes            & \checkmark            & \texttimes            \\ \hline
\textbf{Transaction Speed}       & Moderate (30 TPS)       & High (1000+ TPS)           & High (1000+ TPS)      & High (4000+ TPS)      & High (2000+ TPS)      & Medium (40 TPS)       & Medium (100 TPS)      \\ \hline
\textbf{Cost Efficiency}         & \texttimes              & \checkmark                 & \checkmark            & \checkmark            & \checkmark            & \checkmark            & \checkmark            \\ \hline
\textbf{Data Privacy}            & \texttimes              & \checkmark                 & \texttimes            & \texttimes            & \checkmark            & \texttimes            & \checkmark            \\ \hline
\textbf{Immutability}            & \checkmark              & Configurable               & \checkmark            & \checkmark            & Configurable          & \checkmark            & Configurable          \\ \hline
\textbf{Scalability}             & Moderate                & High                       & High                  & High                  & High                  & Medium                & High                  \\ \hline
\textbf{Interoperability}        & \checkmark              & \texttimes                 & \texttimes            & \checkmark            & \texttimes            & \checkmark            & \texttimes            \\ \hline
\textbf{Ecosystem Maturity}      & \checkmark              & \texttimes                 & \texttimes            & \texttimes            & \texttimes            & Medium                & \texttimes            \\ \hline
\textbf{Community Support}       & \checkmark              & \texttimes                 & \texttimes            & \texttimes            & \texttimes            & \checkmark            & \texttimes            \\ \hline
\end{tabular}%
}
\label{t100}
\end{table*}

After evaluating the features, advantages, and suitability of various blockchain platforms, we have chosen Ethereum as the foundational technology for our digital certification system. Its combination of decentralization, immutability, smart contract functionality, integration with IPFS, and a strong developer ecosystem provides the optimal environment for building a secure, scalable, and reliable digital certification platform.
\subsection{Smart Contracts Design}

\subsubsection{Smart contract Authority of Moroccan Certificate Authority}

\begin{figure*}[htbp]
\begin{center}
\includegraphics[scale=0.7]{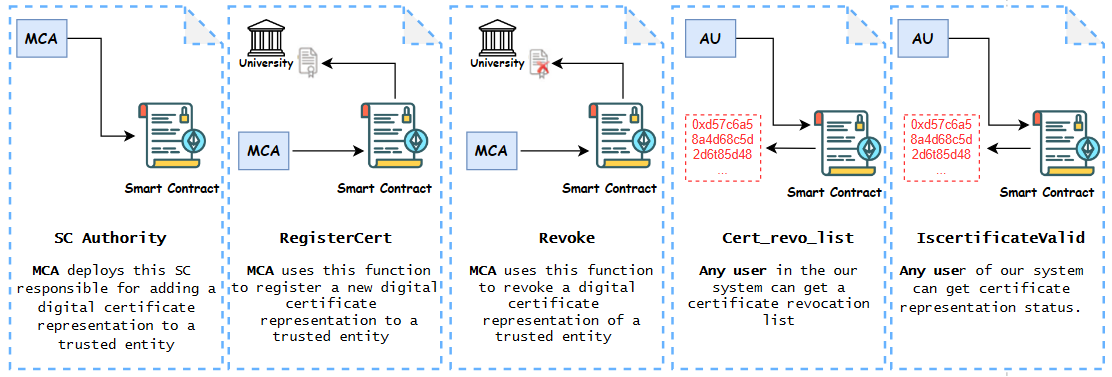} 
\end{center}
\caption{The workflow of the smart contract Authority along with the actors involved in Ethereum blockchain.}
\label{figauth}
\end{figure*}

The smart contract MC\_Authority is responsible for managing the approval and authorization of trusted entities (e.g., universities) to participate in the proposed network. Figure \ref{figauth} illustrates the workflow of this smart contract, highlighting the interactions between different actors within the Ethereum blockchain.

In Figure \ref{figauth}, rectangles marked with “MCA” in steps 1, 2, and 3 denote actions taken by the Moroccan Certificate Authority, while “AU” in steps 4 and 5 represents actions that can be performed by any user within the blockchain network.

The \texttt{MC\_Authority} smart contract is initially deployed on the Ethereum blockchain by MCA. Once deployed and confirmed, MCA can interact with the contract using its unique address. To register a new digital certificate, MCA invokes the \texttt{RegisterCert} function, providing the university’s address, public key, and certificate expiration date. This function can only be called by MCA. Additionally, MCA can revoke certificates using the \texttt{revoke} function as needed.

Internally, the smart contract \textit{MC\_Authority} utilizes the \textit{PKI\_Certificate} struct (see Listing \ref{list1}) to maintain information about each registered certificate. The struct includes fields such as \textit{identity}, \textit{publicKey}, \textit{expiry}, \textit{revoked}, and \textit{registered}, which are mapped to the respective Ethereum addresses of the entities being certified.

\begin{lstlisting}[caption={The Struct PKI\_Certificate}, captionpos=b, frame=single, basicstyle=\scriptsize, label=list1]
struct PKI_Certificate {
        address identity;
        bytes32 publicKey;
        uint expiry;
        bool revoked;
        bool registered;
} 
\end{lstlisting}

The contract also utilizes events to log every addition or modification to the blockchain, providing transparency and traceability. Listing \ref{list2} shows an example of the \textit{Certified} event log generated whenever MCA successfully registers a new certificate. This log contains details such as the sender’s and recipient’s addresses, and the timestamp of the event, making it accessible for any listener applications to monitor the status of certificates on the blockchain.

\begin{lstlisting}[caption={Example of Certified Event Log}, captionpos=b, frame=single, basicstyle=\scriptsize , label=list2]
event Certified(address from, address to, uint date);
logs[
{
 "from": "0xd9145CCE52D386f254917e481eB44e9943F39138",
 "event": "Certified",
 "args":{
 "from": "0x5B38Da6a701c568545dCfcB03FcB875f56beddC4",
 "to": "0xAb8483F64d9C6d1EcF9b849Ae677dD3315835cb2",
 "date": "1659815412"
	}
}]
\end{lstlisting}

The smart contract also includes two public functions: \textit{isCertificateValid(address university)} and \textit{Crt\_revo\_List()}. These functions can be called by any user on the blockchain to verify the status of registered certificates. Specifically:

\begin{itemize}
    \item The \textit{isCertificateValid(address university)} function returns a Boolean value indicating whether a certificate is still valid (\texttt{true}) or has been expired/revoked (\texttt{false}).
    
    \item The \textit{Crt\_revo\_List()} function provides an array of Ethereum addresses corresponding to the entities whose certificates have been revoked. This enables easy verification and tracking of any changes to the status of trusted entities within the network.
\end{itemize}

By utilizing these smart contract functionalities, the \textit{MC\_Authority} ensures a transparent and secure environment for managing digital certificates on the Ethereum blockchain.

\subsubsection{Smart contract university of universities}
\begin{figure*}[htbp]
\begin{center}
\includegraphics[scale=0.75]{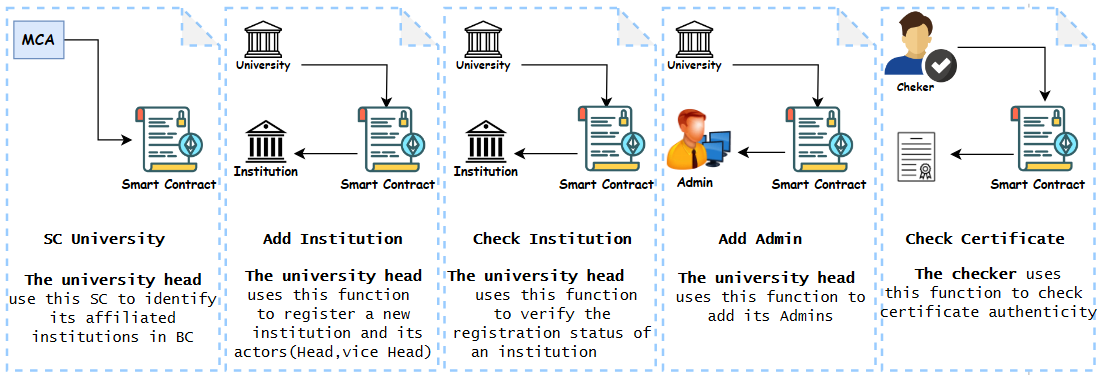}

\end{center}
\caption{The workflow of the smart contract University  along with the actors involved in Ethereum BC.}
\label{figuniv}
\end{figure*}

A university is the second SC published in our workflow. The university constructor defines the trusted entity (e.g. university) as the owner of SC by using the address of the Authority contract previously published, and the address of the trusted entity. It provides each university allowed by MCA to identify its affiliated institutions in BC. Figure \ref{figuniv} shows the functions in the SC along with the actors involved

The digital certificate of the contract owner (university) must be valid to execute the functions in SC. The University contract maintains three primary mappings (see Listing \ref{list3}). The first mapping is dedicated to managing records of its affiliated institutions, ensuring each institution is properly registered and tracked within the system. The remaining two mappings are specifically designed for storing academic certificates: one mapping handles student certificates, and the other manages professor certificates.

\begin{lstlisting}[caption={University's maps}, captionpos=b, frame=single, basicstyle=\scriptsize, label=list3]
 mapping(address => Institution) private Institutions;
  mapping(address => bytes32) public studentCertificates;
  mapping(address => bytes32) public professorCertificates;
 
\end{lstlisting}
The university can invoke the \textit{Add Institution} function to register a new institution to the mapping. The university uses the institution's address as well as the addresses of the head and vice head for inclusion in the mapping. Additionally, the \textit{Check Institution} is used to verify whether an institution has already been registered. If this is the case, the system generates a log indicating 'Institution already exists'.
Administrators can be added to the system, expanding the university’s capacity to manage academic certifications and institutions. The \textit{Add admin} function allows the university (owner) to add an administrator to the system. The event AdministratorAdded is triggered upon adding a new admin.
The certificates issued within the system can be validated through the blockchain, ensuring authenticity and preventing fraud.
The Checker can invoke the \textit{Check Certificate} function to check the validity of a given diploma by referencing the Certifs\_Students contract deployed for the diploma of student and the Certifs\_Profs contract deployed for the diploma of professor. It indicates whether the diploma is valid.

\subsubsection{Smart contract Institution  of universities}

\begin{figure*}[htbp]
\begin{center}
\includegraphics[scale=0.75]{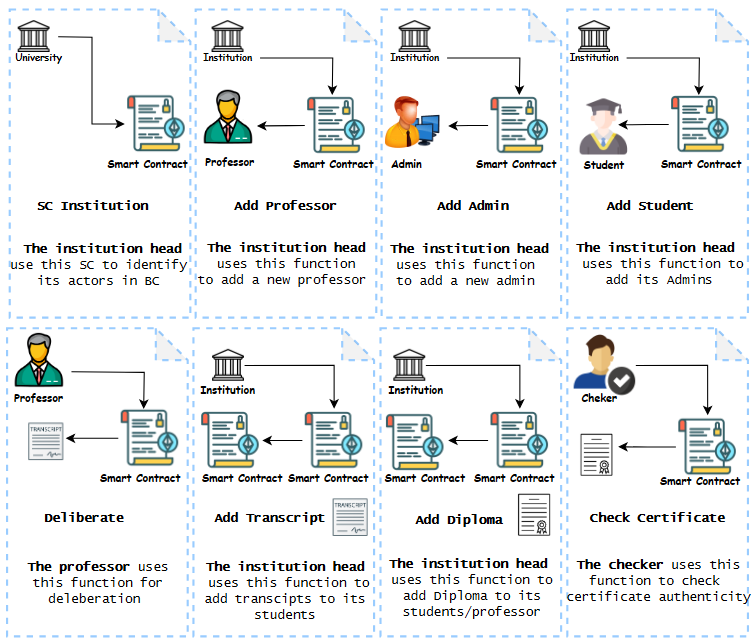}

\end{center}
\caption{The workflow of the smart contract institution along with the actors involved in Ethereum BC.}
\label{figinst}
\end{figure*}

An institution is the third SC published in our workflow. The institution constructor defines the institution (Dean) as the owner of SC by using the address of the university contract previously published, the address of the institution, and the address of the head. It provides for each institution approved by the university to add its actors (authorized administrators, professors, and students) and record transcripts and academic degree certificates for its students and other certificates for its professors. Figure \ref{figinst} shows the functions in the SC along with the actors involved.

Indeed, the contract owner requires validation from the university to execute functions within the smart contract (SC). The institution contract oversees six mappings (refer to Listing \ref{list4}). The first three mappings contain records for professors, administrators, and students. The fourth and fifth mappings are designated for storing transcripts and diplomas. The final mapping is for adding authorized administrators.
\vspace{0.2cm}

\begin{lstlisting}[caption={Institution's maps}, captionpos=b, frame=single, basicstyle=\scriptsize, label=list4]
mapping(address => Professor) private Professors;  
mapping(address => Admin) private Admins;
mapping(address => Student) private students;  
mapping(bytes32 => Transcript) trans; 
mapping(bytes32 => Acdiploma) diploms;
mapping(address => bool) public authorized;
\end{lstlisting}

The Institution smart contract can invoke the \textit{addAuthorized} function to add new authorized administrators and use the \textit{removeAuthorized} to remove authorized administrators. Indeed, the \textit{Add Professor}, \textit{Add Admin}, and \textit{Add Student} functions to register new entities such as professors, admins, and students to the mapping are facilitated through the invocation of the \textit{Add Professor}, \textit{Add Admin}, and \textit{Add Student} functions, respectively. In addition, to register a new transcript for a student, the institution uses the \textit{Add Transcript} function. More precisely, this function checks if the transcript has already been registered; if that is the case, we get a log ‘’Transcript already exists‘’. If not, an internal function to 'SC-Institution' is executed (see Listing \ref{list5}), creating a new instance of SC TranscriptCert.
\begin{lstlisting}[caption={Issuance of a new instance SC TranscriptCert}, captionpos=b, frame=single, basicstyle=\scriptsize,label=list5]
function issueTranscript(address student) 
internal onlyOwner {
       require(msg.sender == owner);
       
 TranscriptCert TrCert = new TranscriptCert(student,owner);
 TrCert.emitt(); }
\end{lstlisting}
In fact, each degree certificate has a certain number of semesters. For example, to get a degree Certificate in General University Studies (Associate's Degree), students must validate four semesters (S1, S2, S3, and S4).
Therefore, the \textit{Add Diploma} function verifies whether the student has been granted the number of semesters (validated transcripts) required for each degree certificate. If that is the case, an internal function to SC Institution is executed (see Listing \ref{list6}), creating a new instance of SC Certs\_Students; if not, we get a log for each degree certification; for example, for Associate's Degree Certification, we get the message "The student has not validated the Diploma of General University Studies.

\begin{lstlisting}[caption={Issuance of a new instance SC Academic Certs\_Students}, captionpos=b, frame=single, basicstyle=\scriptsize,label=list6]
function issueDiploma(address university, address student)
internal {require(Autho.isCertificateValid(university));
 Certs_Students diploma = new Certs_Students(student,
 university,owner);
 diploma.emitt();
    }
\end{lstlisting}
For the professor certificate, an internal function of the SC Institution is executed (see Listing \ref{list7}), creating a new instance of SC Certs\_Profs.
\begin{lstlisting}[caption={Issuance of a new instance SC Academic Certs\_Profs}, captionpos=b, frame=single, basicstyle=\scriptsize,label=list7]
function issueDiploma(address university, address student)
internal {require(Autho.isCertificateValid(university));
 Certs_Profs diploma = new Certs_Profs(student,
 university,owner);
 diploma.emitt();
    }
\end{lstlisting}

The \textit{deleberate} function allows a professor to deliberate and update a student’s grade in a secure, transparent manner while maintaining academic integrity and accountability.

The Checker can invoke the \textit{Check Certificate} function to check the validity of a given diploma by referencing the Certifs\_Students contract deployed for the diploma of the student and the Certifs\_Profs contract deployed for the diploma of the professor. It indicates whether the diploma is valid.

Following a successful transaction, the certificate hash is recorded on the Ethereum blockchain, while a pre-defined template for a specific academic certificate is saved in the student portfolio via the IPFS network, facilitated by the storage smart contract.

\subsubsection{Smart contract Certifs\_Students }

When a student successfully completes a degree, the authorized university institution creates a new instance of the SC Certifs\_Students contract. This contract takes the student's address, the trusted entity's address (as issuer\_president), and the institution's address (as issuer\_dean) as inputs. The emit function in SC Certifs\_Students records the current timestamp as the date the student completes their degree. Diploma and degree certification data are stored in the form of the Diploma struct, as shown in Listing \ref{list8}.

\begin{lstlisting}[caption={The struct Academic Diploma}, captionpos=b, frame=single, basicstyle=\scriptsize,label=list8]

    struct Acdiploma {
        address issuer_dean;
        address issuer_presedent;
        address receiver;
        bytes32 Degre;
        uint date;
        uint note;
        bool status;
    }
\end{lstlisting}

In fact, the trusted entity can call the \textit{revoke} function in SC Certifs\_Students to refuse a Diploma Instance when we have an error in student enrollment by using a special function, which is \textit{selfdestruct}; this function definitively deletes an SC by clearing its code and data.

\subsubsection{Smart contract TranscriptCert }

The trusted entity creates a new instance of the SC-TranscriptCert contract to issue a TranscriptCert to a student. This contract takes the student's address and the trusted entity's address (as issuer\_dean) as inputs. The emit function in TranscriptCert records the current timestamp as the date when the student receives their transcript. Transcript data is stored in the form of the TranscriptCert struct. as shown in Listing \ref{list9}.

\begin{lstlisting}[caption={The struct Student TranscriptCer}, captionpos=b, frame=single, basicstyle=\scriptsize, label=list9]

      struct TranscriptCert {
        address issuer;
        address receiver;
        bytes32 Degre;
        bytes32 Semstre; 
        uint score;  
        uint date;
        uint note;
        bool status;
    }
\end{lstlisting}

In fact, the trusted entity can use the same function in SC Certifs\_Students to refuse a diploma instance.

\subsubsection{Smart contract Certifs\_Profs }

The trusted entity creates a new instance of the SC Certifs\_Profs contract to issue a certificate to a professor. This contract takes the professor’s address, the trusted entity's address (as issuer\_president), and the institution’s address (as issuer\_dean) as inputs. The emit function in Certifs\_Profs records the current timestamp as the date when the professor receives their certificate. Diploma and certification data are stored in the form of the PrDiploma structure, as presented in Listing \ref{list10}.

\begin{lstlisting}[caption={The struct Professor Diploma}, captionpos=b, frame=single, basicstyle=\scriptsize, label=list10]

    struct PrDiploma{
        address issuer_head;
        address issuer_Institution ;
        address receiver;
        bytes32 Degre;
        bytes32 subject;
        bytes32 Department;
        uint period;
        bool status;
        uint date;
    }
\end{lstlisting}

In fact, the trusted entity can use the same function in SC Certifs\_Students to refuse a diploma instance.

\subsubsection{Smart contract storage of Universities}

\begin{figure*}[htbp]
  \centering
    \includegraphics[scale=0.75]{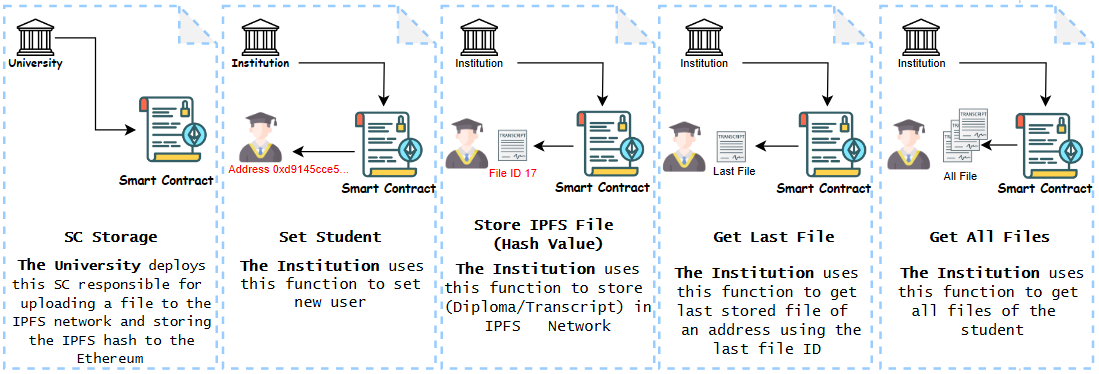}
  \caption{  {The workflow of the smart contract storage along with the actors involved in Ethereum BC.}}
  \label{figstt}
\end{figure*}

The trusted entity creates a new instance of SC Storage. In fact, it can call the store function to upload a file to the IPFS network and add a Content Identifier, or CID (cryptographic hash of file), to the Ethereum network. More precisely, store the IPFS hash in an Ethereum contract that associates the user's Ethereum account with the IPFS file hash. The SC Storage uses a set of functions, which are:
\begin{itemize}
\item setUser(address \_add) : set user 
\item getLastUserID (address \_add): get last file id
\item getLast (address \_add): get the last stored file of an address using the last file id 
\item getAll(): get all files of the user :
\end{itemize}
Figure \ref{figstt} shows the functions in the SC along with the actors involved. 

In fact, the contract Storage maintains tree mappings (see Listing \ref{list11}). The first
used to link an address with the stored files; second used to check if an address already stores some files; third used the address to track the number of stored files. \vspace{0.2cm}
\begin{lstlisting}[caption={Storage's maps}, captionpos=b, frame=single, basicstyle=\scriptsize,label=list11]
mapping(address => mapping(uint256 => UserFile)) 
public UserFiles;
mapping(address => bool) public isSet;
mapping(address => uint256) public lastID;
\end{lstlisting}

\section{Implementation Results and Discussion} 
\label{sec5} 
In this section, we elaborate on the integration of smart contracts into our proposed system. 

\begin{figure*}[htbp]
  \centering
    \includegraphics[scale=0.6]{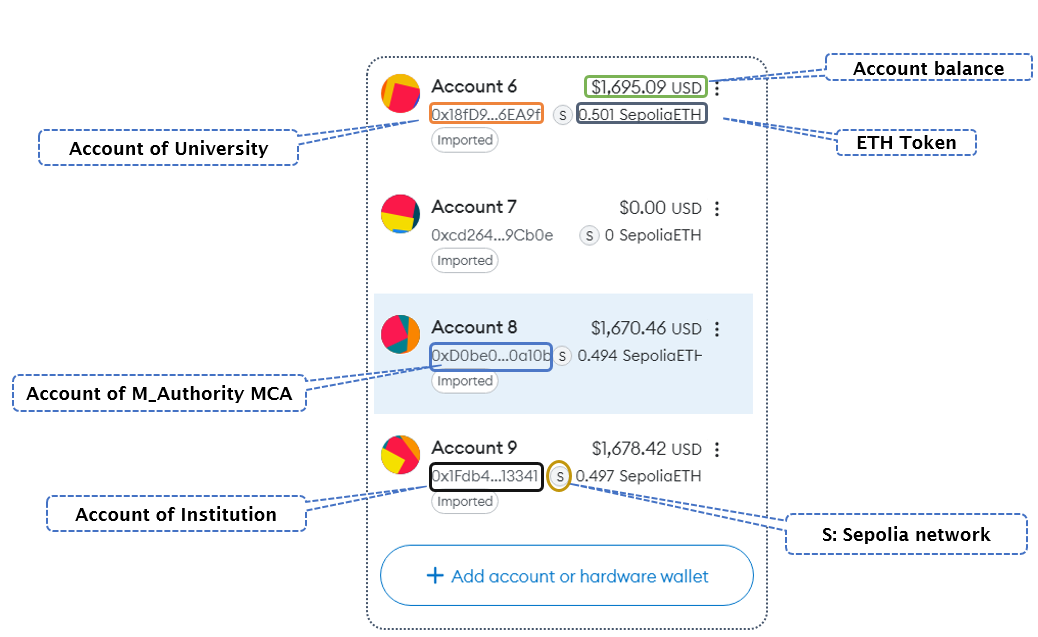}
  \centering 
     \caption{{
     Account balance after adding the Sepolia network and obtaining ETH funds in the MetaMask wallet.}}
  \label{fig9c}
\end{figure*}

Figure \ref{fig9c} displays the account balance. Figures and graphs offer visual insights into the test transactions and other pertinent data.

\subsection{Implementation setup}
The implementation of the BlockMEDC system was carried out on the Sepolia Ethereum test network using the Remix IDE and Solidity for smart contract development. The setup began by configuring the MetaMask wallet and connecting it to the Sepolia network, where network parameters such as the RPC URL, Chain ID, and SepoliaETH faucet were established. This enabled seamless interaction with the blockchain during testing and deployment.
\begin{algorithm}[H]
\caption{Implementation setup for our system.}
\begin{algorithmic}[1]
    \State \textbf{Initialize Metamask and Connect to sepolia Network}
    \State Set network parameters:
    \begin{itemize}
        \item Network Name: \texttt{Sepolia test network}
        \item RPC URL: \url{https://sepolia.infura.io/v3/}
        \item Chain ID: \texttt{11155111}
        \item Currency Symbol: \texttt{SepoliaETH}
        \item Explorer URL: \url{https://sepolia.etherscan.io}
    \end{itemize}
    \State 
   Use a faucet to fund our Sepolia testnet account and transfer test ETH to our MetaMask wallet.

    \State \textbf{Deploy Smart Contracts on sepolia Network}
    \State Open Remix IDE, compile Solidity smart contracts, and deploy them to the Sepolia network.
   \State Configure smart contracts for seamless operation with the sepolia network.
   \State Use Python libraries to connect to the sepolia network.
   \State Use Python to interact with the smart contracts.
    \State \textbf{Run all Smart Contracts}
    \State Get data: certificates, transcripts, student diplomas, professor diplomas.
    \State Analyze results to evaluate the effectiveness of smart contracts for securing educational digital certificates.
\end{algorithmic}
\label{algo1}
\end{algorithm}

Algorithm \ref{algo1} illustrates the step-by-step procedure followed to implement our system. This algorithm covers the entire implementation pipeline, from initializing the MetaMask wallet to interacting with the deployed smart contracts. After deploying the contracts, Python (Version 3.8) scripts were used to interact with the smart contracts, enabling operations such as certificate issuance, management, and verification within the system.
The smart contracts, including MC\_Authority, University, Institution, and Storage, were deployed using the Remix IDE (Version 0.22.2) on the Sepolia network. The system leveraged Infura’s RPC nodes to facilitate remote interactions and IPFS for decentralized off-chain storage of academic documents, ensuring efficient and scalable data management. MetaMask, a trusted Ethereum wallet, was used to fund test accounts and execute transactions on the blockchain, while the SepoliaETH token was used to cover gas fees during contract interactions.
\renewcommand{\arraystretch}{1.2} 
\begin{table}[H]
\centering
\caption{Implementation Details.}
\resizebox{1\columnwidth}{!}{
    \begin{tabular}{|p{3.5cm}|p{3cm}|}  
        \hline
        \textbf{Attribute} & \textbf{Value} \\ \hline
        Ethereum Network & Sepolia \\ \hline
        Integrated development environment (IDE) & Remix IDE (version 0.22.2) \\ \hline
        Smart Contract Programming Language & Solidity (Version 0.4.23) \\ \hline
        Programming language used for interacting with our system & Python (Version 3.8) \\ \hline
        Ethereum wallet & MetaMask \\ \hline
        Ethereum Token & Sepolia ETH \\ \hline
        Ethereum Storage & IPFS \\ \hline
        RPC node & Infura \\ \hline
        Block explorer URL & https://sepolia.etherscan.io/ \\ \hline
    \end{tabular}%
}
\label{tab100}
\end{table}
Table \ref{tab100} presents the details of the tools and technologies utilized in the implementation process. This setup ensured a secure and efficient deployment, enabling real-time testing and validation of smart contracts within the proposed system.

\subsection{Results}
\subsubsection{Smart contract MC\_Authority of MCA}
\begin{figure*}[htbp]
  \centering
    \includegraphics[scale=0.44]{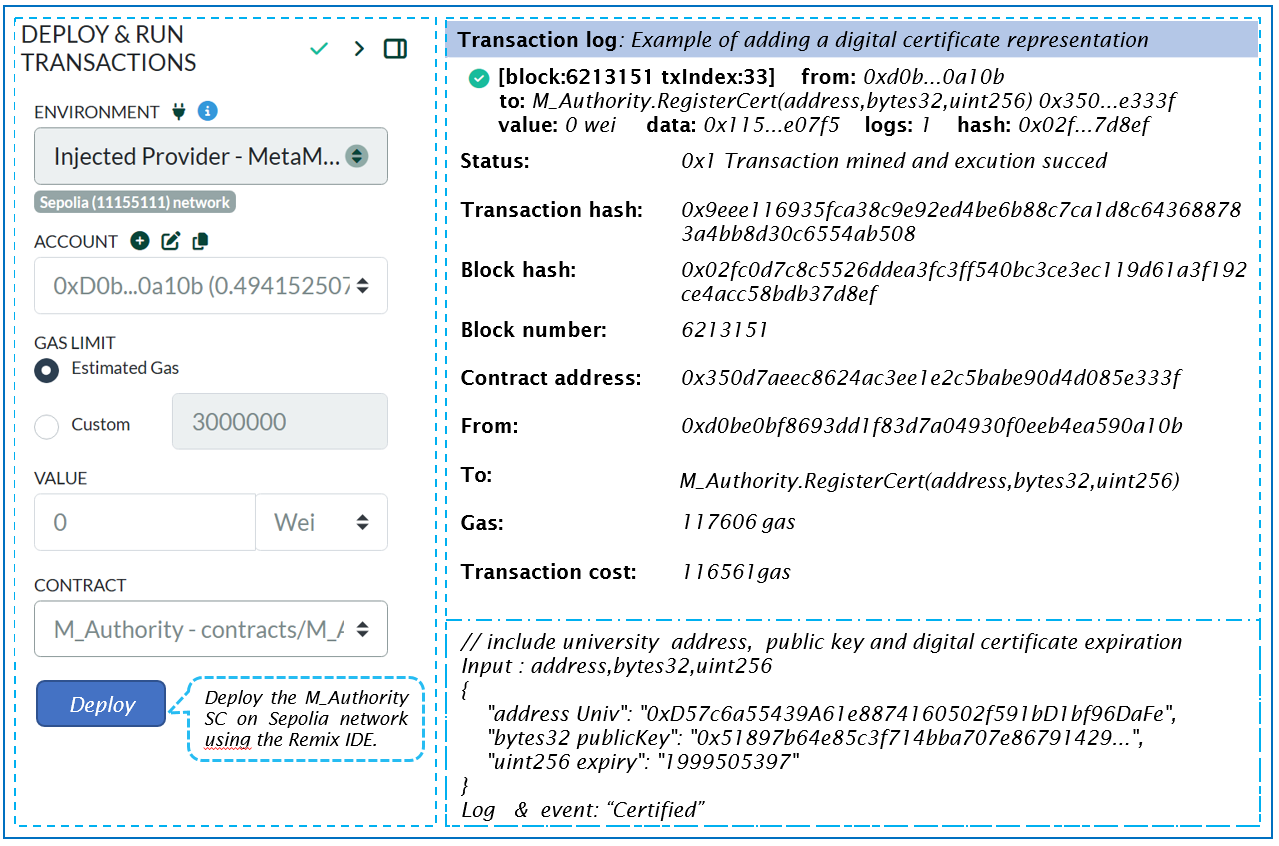}
  \centering 
     \caption{{
     Remix IDE screen of our deployed MC\_Authority smart contract.}}
  \label{fig10}
\end{figure*}

Figure \ref{fig10} presents the transaction details for deploying an instance of the MC\_Authority smart contract. It highlights key elements, such as the transaction status, contract address, sender address, and destination, which refers to the contract's constructor.

Once the smart contract is published on the Ethereum BC, MCA can invoke the contract's functions using the address shown in Figure \ref{fig10}. Specifically, to add a digital certificate representation to the blockchain, MCA calls the \textit{RegisterCert} function.

To verify the authenticity of a university certificate, the \textit{isCertificateValid} function is called. Additionally, the contract allows MCA to revoke a digital certificate using the \textit{revoke} function. A certificate is also automatically marked as invalid upon expiration. The \textit{cert\_revo\_list} function returns a list of addresses associated with entities whose certificates have been revoked, ensuring transparency and traceability of the certification status.

\begin{figure*}[htbp] 
  \centering
  \includegraphics[scale=0.43]{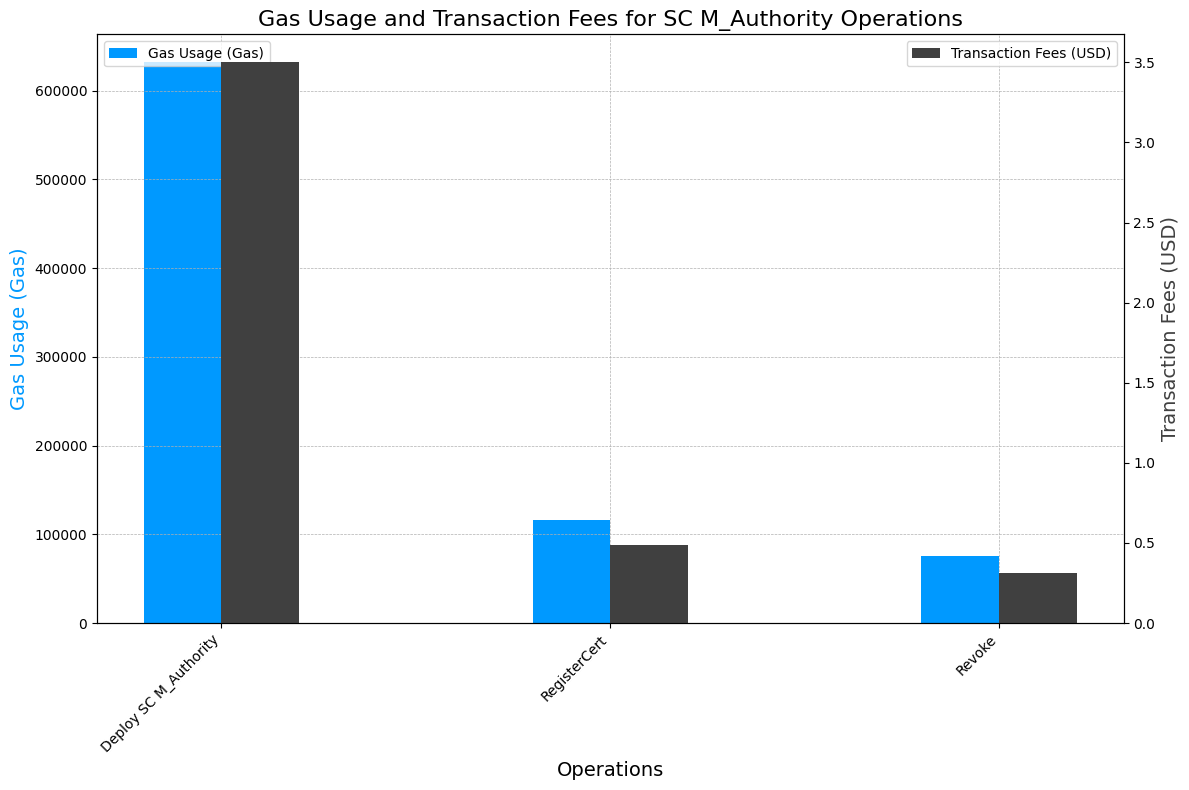}
  \caption{{Transaction Fee screen: Deployment and interaction with a smart contract 'MC\_Authority' using the Sepolia network.}}
  \label{fg14}
\end{figure*}

The breakdown of transaction fees for deploying and interacting with the smart contract MC\_Authority is outlined in Figure \ref{fg14}.
This figure showcases the gas costs related to both deploying the smart contract and executing its functions. It presents the gas costs associated with different operations, such as the initial deployment of the smart contract, the addition of digital certificate representations for universities, and the revocation of such representations.

\subsubsection{Smart contract University of universities}
\begin{figure*}[htbp]
  \centering
    \includegraphics[scale=0.43]{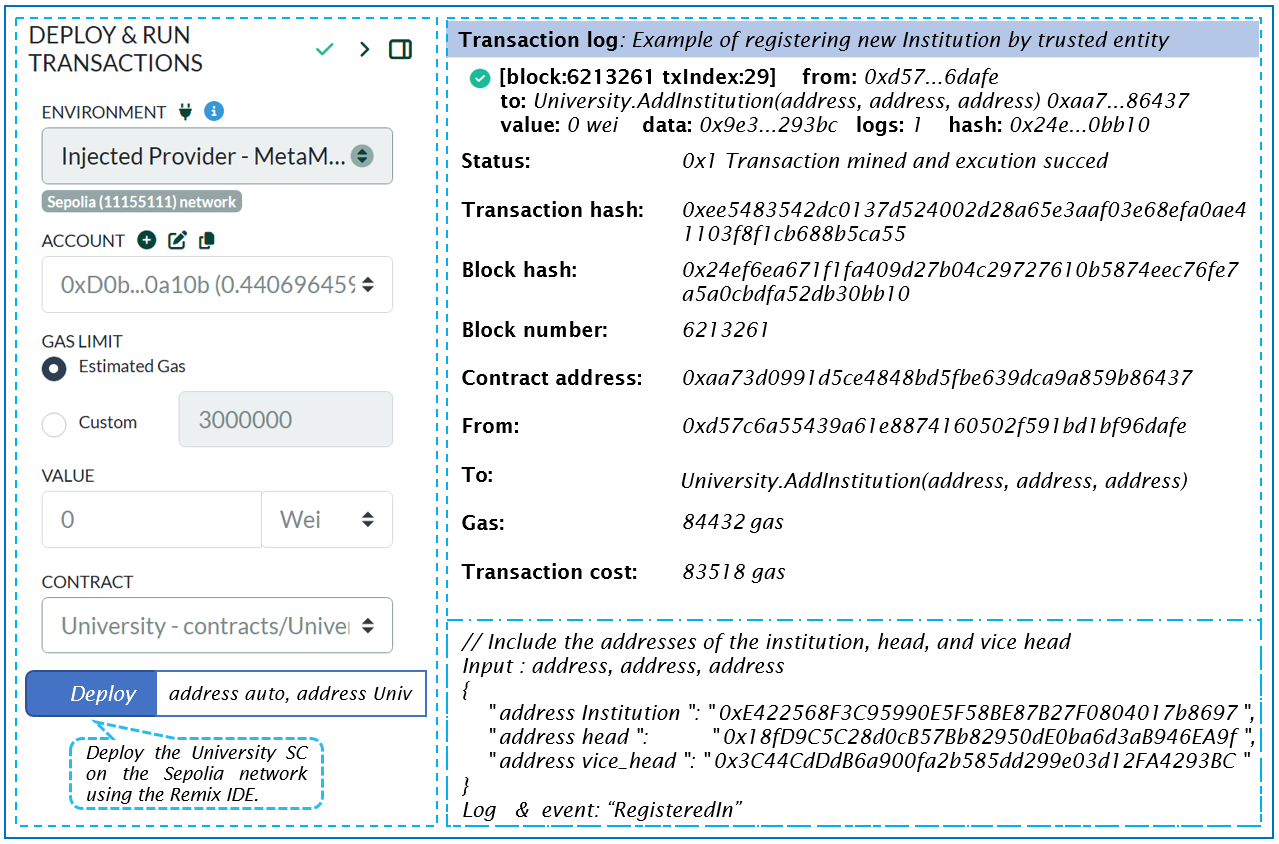}
  \caption{  {Remix the IDE screen of our deployed university smart contract.}}
  \label{fig11}
\end{figure*}

Figure \ref{fig11} displays the transaction details for creating an instance of the university smart contract. It also includes the specific transaction information for adding an affiliated institution.

The process of adding an institution to the blockchain, where ownership is restricted to a university approved by the MCA, is managed through the \textit{AddInstitution()} function. The university head can also assign administrators to the system using the \textit{addAdministrator()} function.

Additionally, the \textit{issueDiplomaP()} function allows the university head to issue academic diplomas for students and professors, secured with a hashed signature (barcode data) for enhanced authenticity.

For verification purposes, the \textit{CheckInstitution()} function confirms the accreditation status of affiliated institutions, while the \textit{validateCertStudent()} and \textit{validateCertProf()} functions enable a checker to validate academic certifications for students and professors, respectively.

\begin{figure*}[htbp] 
  \centering
  \includegraphics[scale=0.43]{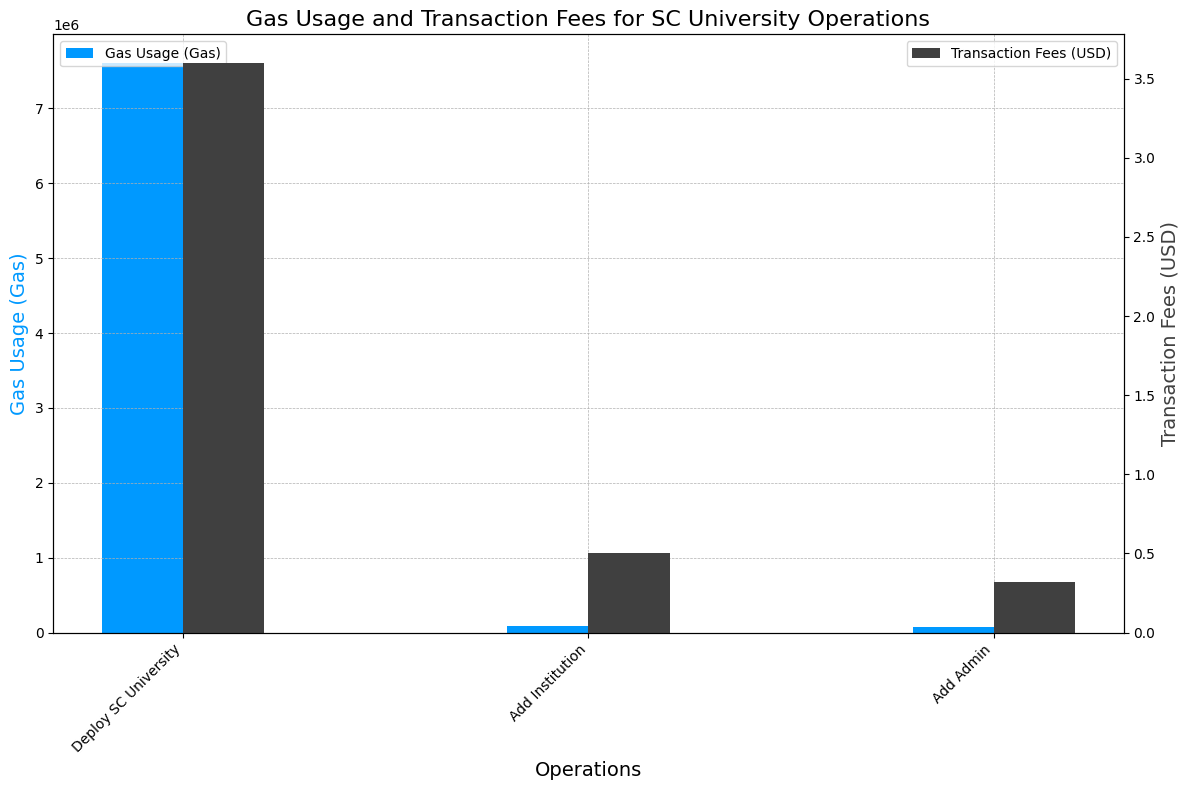}
  \caption{Transaction Fee screen: Deployment and interaction with a smart contract 'University' using the sepolia network}
  \label{fgu1}
\end{figure*}

Figure \ref{fgu1} presents the transaction fee details for deploying and interacting with the university smart contract. The figure illustrates the gas costs associated with deploying the contract and executing its functions. From right to left, it shows the gas consumption for various operations, such as the initial deployment of the smart contract, adding affiliated institutions, and assigning administrators to the system.

\subsubsection{Smart contract Institution of universities}

\begin{figure*}[htbp]
  \centering
    \includegraphics[scale=0.45]{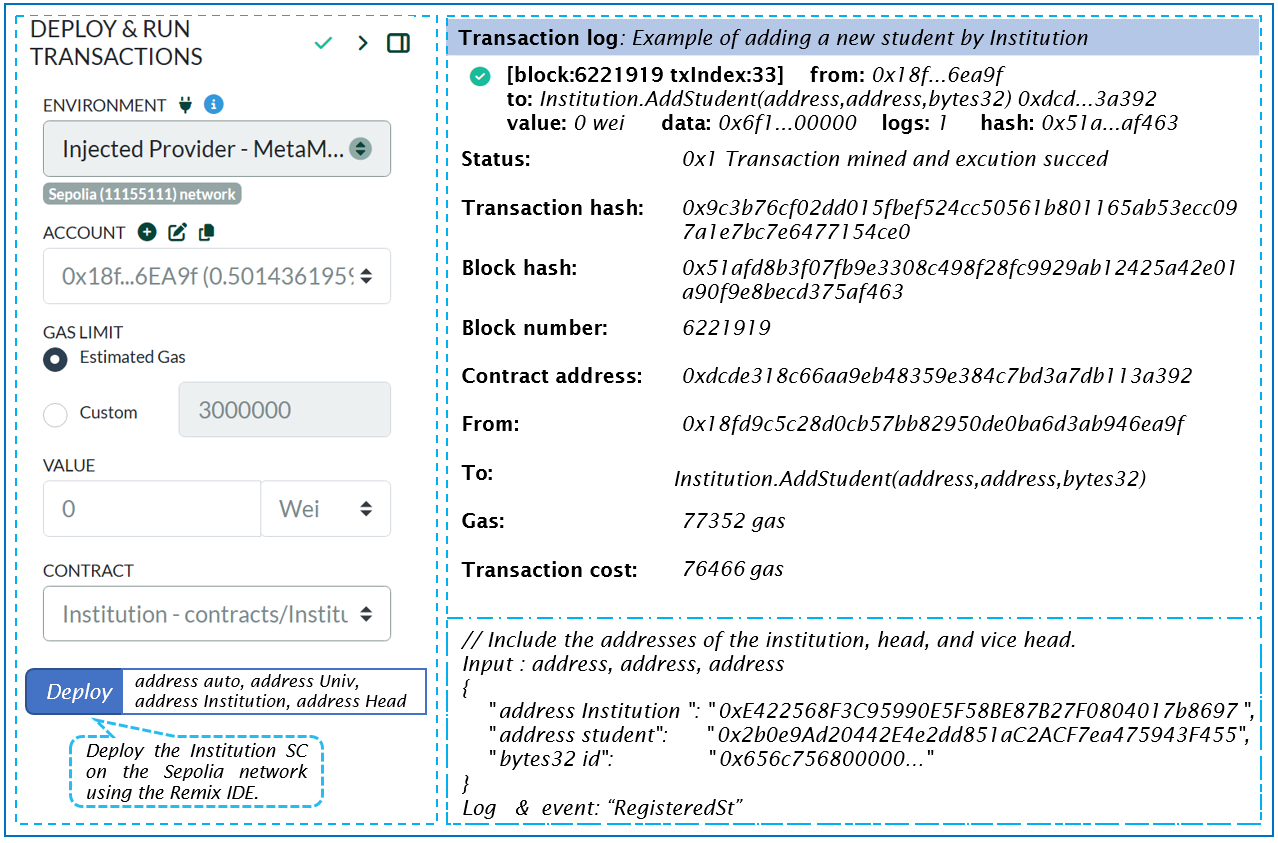}
  \caption{  {Remix IDE screen of our deployed Institution smart contract.}}
  \label{fig12}
\end{figure*}

Figure \ref{fig12} illustrates the transaction details involved in creating an instance of the Institution smart contract. Furthermore, it presents the specific transaction information for adding students to the system.

The institution contract offers various functions to manage authorized entities and academic records. Key functionalities include adding and removing authorized users with the \textit{addAuthorized()} and \textit{removeAuthorized()} functions. It also facilitates the addition of professors, administrators, and students through the \textit{AddProfessor()}, \textit{AddAdmin()}, and \textit{AddStudent()} functions, respectively. Academic achievements, such as transcripts and diplomas, can be recorded using functions like \textit{AddTranscript()}, \textit{issueDiploma()}, \textit{issueTranscript()}, and \textit{issueProfessorDiploma()} for issuing professor diplomas. Additionally, official documents such as doctoral diplomas can be issued through the \textit{issueDocDiploma()} function.

The \textit{Deliberate()} function enables professors to participate in grade deliberations and update student grades.

Users can interact with the contract to verify permissions and retrieve information using functions like \textit{isStudentAuthorized()}, \textit{getStudentInfo()}, \textit{isAdminAuthorized()}, and \textit{isProfessorRegistered()}. The contract also provides methods to verify the existence of academic documents such as transcripts and diplomas through \textit{areStudentTranscriptsExist()} and \textit{isStudentDiplomaExist()}. Finally, the \textit{validateCertStudent()} and \textit{validateCertProf()} functions enable a checker to validate academic certifications for students and professors, respectively.

\begin{figure*}[t] 
  \centering
  \includegraphics[scale=0.43]{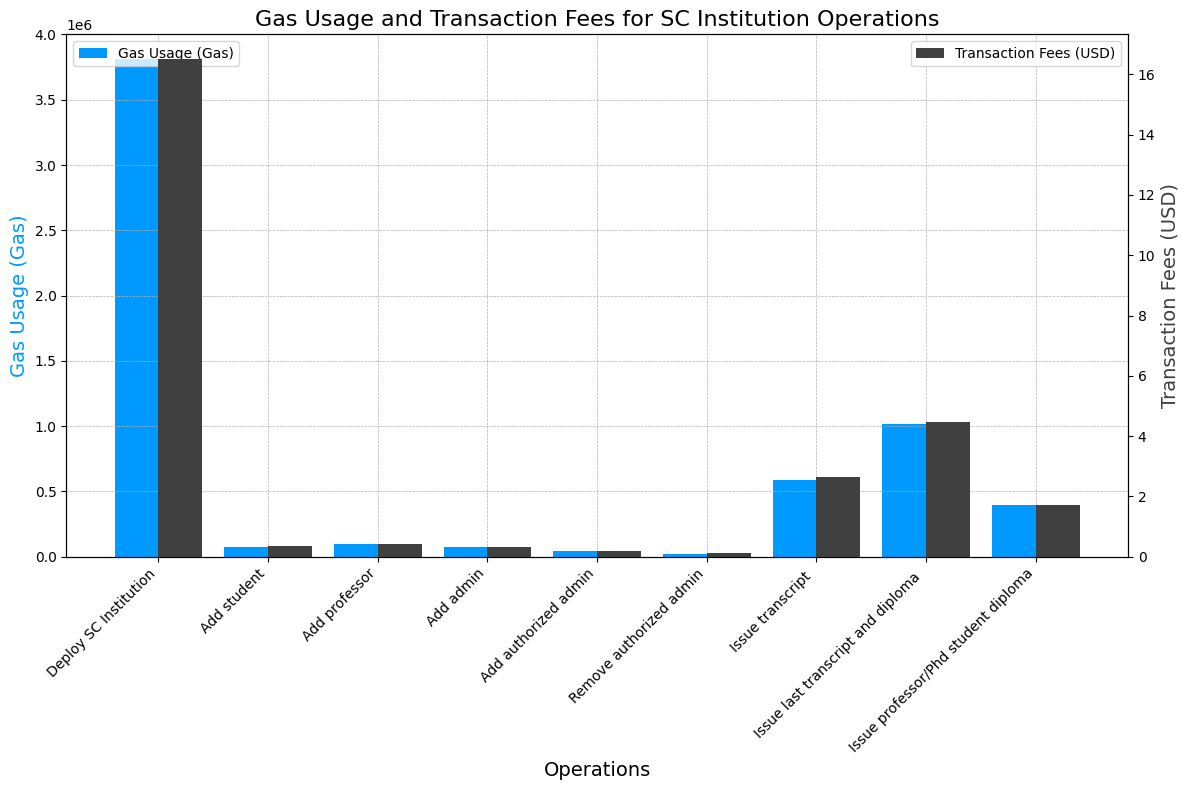} 
  \caption{{Transaction Fee screen: Deployment and interaction with a smart contract 'Institution' using the sepolia network.}}
  \label{fgin}
\end{figure*}

Figure \ref{fgin} illustrates the transaction fees associated with deploying and interacting with the Institution smart contract. It details the gas costs for both the initial deployment and various operational activities. These activities include adding students, professors, and administrators (including vice heads), removing an authorized administrator, adding transcripts, issuing the final transcript for the Associate's Degree certificate, validating four transcripts for diploma issuance, and issuing doctoral, professor, and student diplomas.

\subsubsection{Smart contract storage of data}
\begin{figure*}[t]
  \centering
   \includegraphics[scale=0.43]{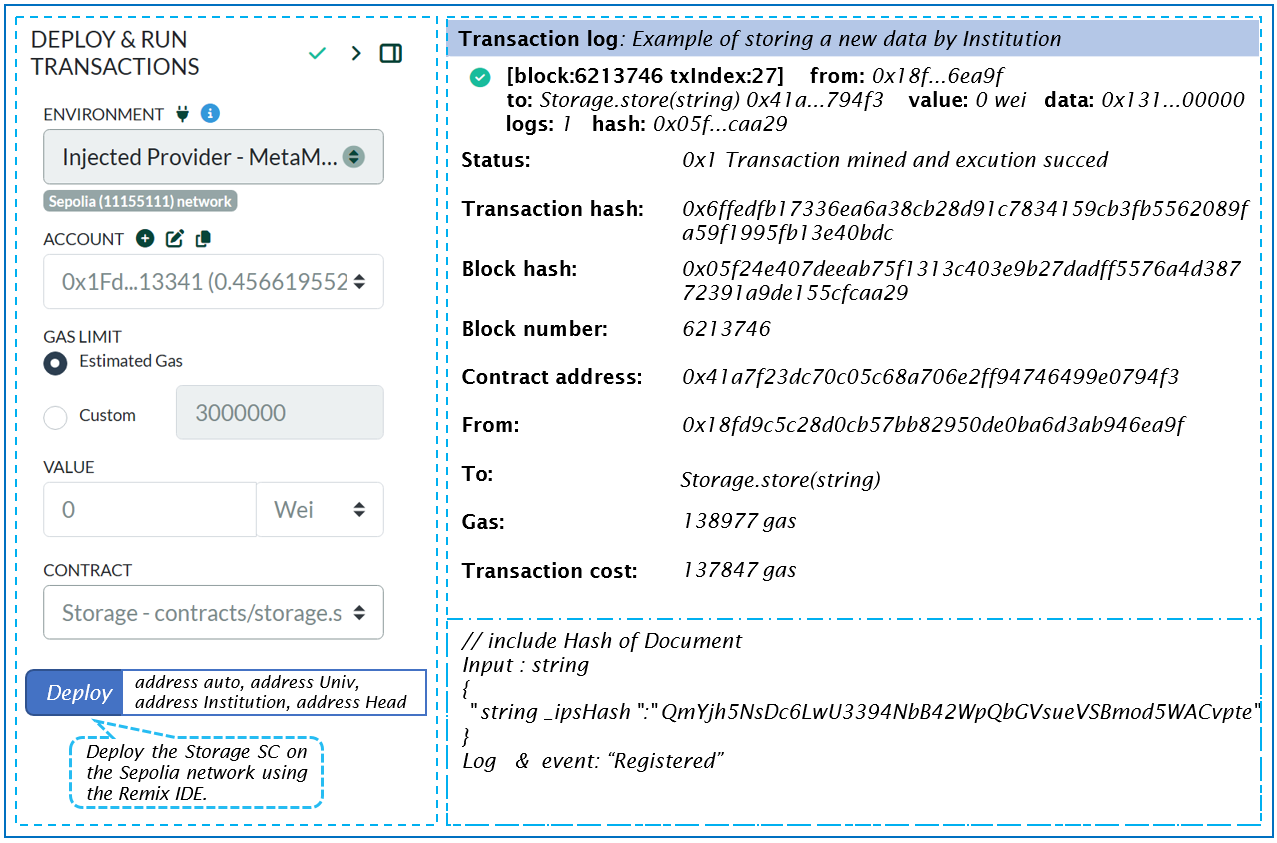}
  \caption{  {Remix IDE screen of our deployed storage smart contract.}}
  \label{fig14}
\end{figure*}

Figure \ref{fig14} illustrates the transaction details for creating an instance of the Storage smart contract. It also details the specific transaction information for storing both student and professor data.

The storage contract provides a variety of functions for handling academic records. These include \textit{setUser()} and \textit{updateUser()}, which enable the addition and modification of user data, and \textit{store()}, which facilitate the storage of academic documents.

Additionally, users interact with the contract through functions such as \textit{getLastUserID()}, \textit{getLast()}, and \textit{getAll()}, which allow retrieval of the user's ID, the latest file associated with the user, and all files stored in the user's IPFS, respectively.

\begin{figure*}[t]
  \centering
  \includegraphics[scale=0.4]{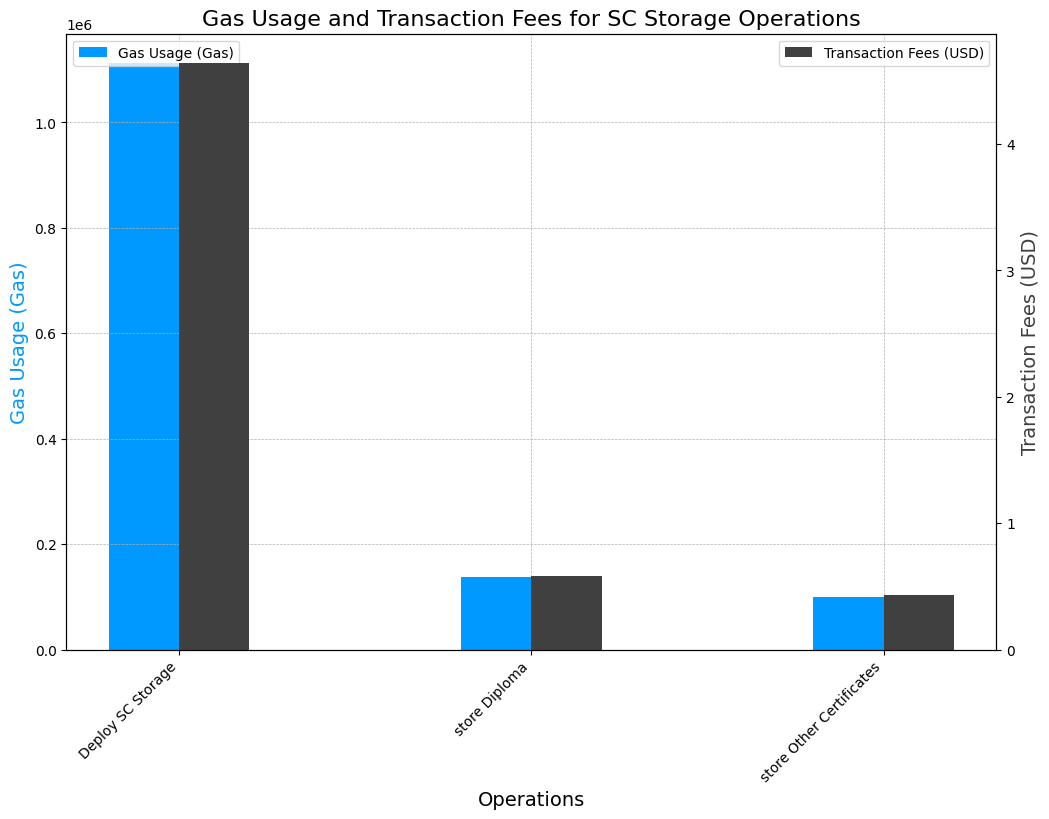}
  \caption{{Transaction Fee Screnn: Deployment and interaction with a smart contract `Storage' using the sepolia network. }}
  \label{fgs2}
\end{figure*}

The breakdown of transaction fees for deploying and interacting with the smart contract storage is detailed in figure \ref{fgs2}. This visualization illustrates the gas costs associated with deploying and executing functions within smart contract storage. Transitioning from right to left on the figure, we discern the gas costs for deploying the SC and storing academic documents for students and professors.

\subsubsection{Discussion}

Table \ref{tabgas} provides an overview of the gas costs associated with issuing digital certificates per student or professor, which are recorded on the Ethereum blockchain and stored in their respective IPFS-based e-portfolios.

In smart contract development, contractors generally incur higher gas costs than other functions for several reasons. First, constructors handle larger input sizes as they initialize state variables and perform crucial setup operations. For example, deploying the smart contract that handles transcripts incurs a gas cost of 590.460 Gwei, approximately 2.64 USD. Similarly, the contract responsible for issuing diplomas costs 413.092 Gwei, or around 1.83 USD. Additionally, constructors allocate internal variables in non-volatile memory, which requires resource-intensive Ethereum Virtual Machine (EVM) instructions, leading to increased gas costs. However, it is important to note that constructors are executed only once during the contract's deployment, meaning these higher costs are a one-time expense.

Transitioning to the discussion of costs per student, the gas expenses include not only registering the student but also recording academic achievements and creating instances of various contract types, such as Certifs\_Students, Certifs\_Profs, and TranscriptCert, as well as storing the related data. For example, registering other types of certificates, such as completion or merit certificates, incurs a gas cost of 394.174 Gwei (1.70 USD). Additionally, storing documents within the IPFS-based e-portfolio, which is essential for the immutability and decentralization of student records, requires a gas cost of 100.434 Gwei, equivalent to 0.43 USD. While constructors may have a higher initial cost, they play a fundamental role in establishing the contract's overall functionality. Meanwhile, the recurring costs per student include a broader range of operations necessary for managing and updating student records and achievements within the smart contract framework.

\renewcommand{\arraystretch}{2.1}
\begin{table}[htbp]
\centering

\caption{Approximation of Gas Costs per Student and Professor}
\begin{tabular}{|l|c|c|}
\hline
\textbf{Description}                          & \textbf{Gas Cost (Gwei)} & \textbf{Cost in USD} \\ \hline
\textbf{Transcript}                           & 590.460                 & 2.64                 \\ \hline
\textbf{Diploma}                      &  413.092                &  1.83 
\\ \hline
\textbf{Other certificate}                    &  394.174                & 1.70                 \\ \hline
\textbf{Storage of a document}        & 100.434                 & 0.43                 \\ \hline
\end{tabular}
\label{tabgas}
\end{table}

\subsection{Limitations and Challenges of the Proposed System}

While the current version of the BlockMEDC system demonstrates promising potential, it also presents certain limitations that need to be addressed:

\begin{itemize} \item The BlockMEDC system is currently designed to serve higher education institutions exclusively and does not yet support other levels of the Moroccan education system. \item The cost associated with each student remains relatively high. \item As BlockMEDC is built on a public blockchain, Ethereum, there are potential privacy concerns that need to be considered and mitigated. \end{itemize}

\section{Conclusion and Future Work}
\label{sec6}

This paper introduced BlockMEDC, a blockchain-based system designed to secure and streamline the management of Moroccan educational digital certificates. Leveraging Ethereum smart contracts and the IPFS network. The BlockMEDC system provides authenticity, transparency, and tamper resistance, addressing key challenges of traditional digital certification methods. Although the current implementation focuses on higher education.
Future directions include optimizing transaction costs using Layer-2 solutions, enhancing privacy through zero-knowledge proofs, and extending support to other educational levels, such as primary, secondary, and vocational education. Additional future work will focus on integrating machine learning for fraud detection, improving interoperability with international educational platforms, and exploring the use of decentralized identity frameworks for secure user management. These enhancements will ensure that BlockMEDC evolves into a comprehensive and scalable solution that aligns with the broader goals of digital transformation in education.

\section*{Appendices} 

The following resources provide additional information on the implementation and testing of the proposed system:

\begin{itemize}
    \item \textbf{Implementation:} The complete implementation of our proposed digital Moroccan education system, based on blockchain smart contracts, can be found at the following link:  
    \url{https://github.com/MedFartitchou/BlockMEDC}
    
    \item \textbf{Testing:} The system was tested using the Python programming language to interact with the smart contracts. Detailed testing procedures and code are available at:  
    \url{https://shorturl.at/rkcHg}
\end{itemize}


\begin{IEEEbiography}[{\includegraphics[width=1in,height=1.25in,clip,keepaspectratio]{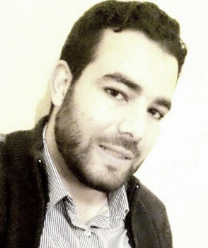}}]{Mohamed Fartitchou} received his Master’s degree in Networks and Intelligent Systems from the Faculty of Sciences and Technology of Fez (FST-Fès), USMBA, in 2015. Since 2016, he has been a high school mathematics teacher. Currently, he is pursuing his Ph.D. in Computer Science at the University Mohammed Premier, Faculty of Sciences, Oujda, Morocco. His research focuses on cybersecurity and the Internet of Things (IoT), with more than six publications in international journals and conferences.
\end{IEEEbiography}
\begin{IEEEbiography}[{\includegraphics[width=1in,height=1.5in,clip,keepaspectratio]{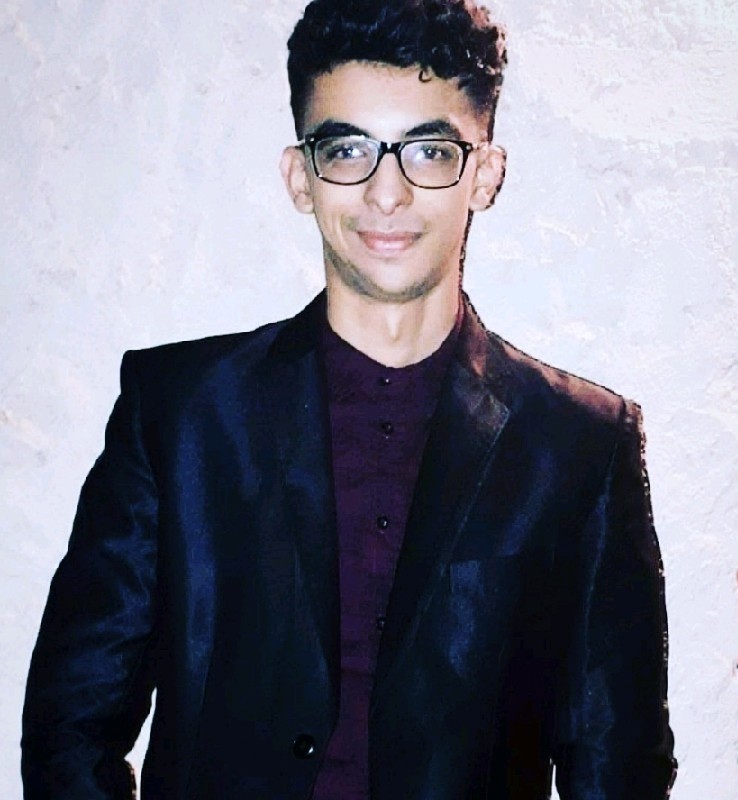}}]{ISMAIL LAMAAKAL}(Graduate Student Member, IEEE) received a Master of Science degree in Computer Science from the Multidisciplinary Faculty of Nador, University Mohammed Premier, Oujda, Morocco. He is currently advancing towards his Ph.D. in Computer Science at the same esteemed institution. As an Artificial Intelligence Scientist, his research primarily focuses on the innovative integration of Tiny Machine Learning, the Internet of Things (IoT), Embedded Systems and Cybersecurity. His work is characterized by its pioneering approach in the field, emphasizing practical applications and advancements in these interconnected domains. His contributions are marked by a commitment to pushing the boundaries of AI and its applications in the modern technological landscape.
\end{IEEEbiography}
\begin{IEEEbiography}[{\includegraphics[width=1in,height=1.35in,clip,keepaspectratio]{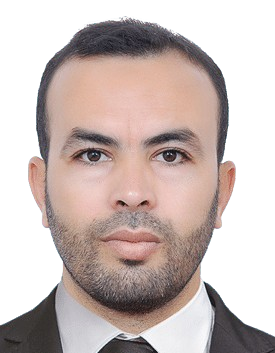}}]{Khalid El Makkaoui} received his master’s degree in networks and systems and a Ph.D. in computer science from Hassan 1st University, Settat, Morocco, in 2014 and 2018. Since 2019, he has been an Associate Professor with the Department of Computer Science at the Multidisciplinary Faculty of Nador, University Mohammed Premier, Oujda, Morocco. His research interests focus on cybersecurity and artificial intelligence. He has published over 50 papers (book chapters, international journals, and conferences).
\end{IEEEbiography}

\begin{IEEEbiography}[{\includegraphics[width=1in,height=1.35in,clip,keepaspectratio]{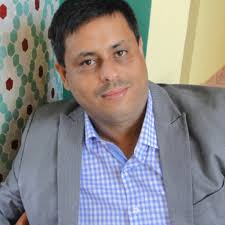}}]{Zakaria El Allali}  received the B.Sc. degree in Operational Research and Statistics and the Ph.D. degree in Mathematics from Mohamed I University, Oujda, Morocco, in 1995 and 2000, respectively. Since 2007, he has been a researcher and a professor with the Polydisciplinary Faculty of Nador, Mohamed I University, Oujda, Morocco. His research interests include mathematical modeling and its applications in engineering and quantum sciences, cybersecurity and artificial intelligence.
\end{IEEEbiography}

\begin{IEEEbiography}[{\includegraphics[width=1in,height=1.25in,clip,keepaspectratio]{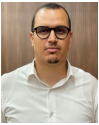}}]{Yassine Maleh} is a professor of cybersecurity and IT governance at Sultan Moulay Slimane University, Morocco. He is the founding chair of IEEE Consultant Network Morocco and founding president of the African Research Center of Information Technology \& Cybersecurity. He is a senior member of IEEE and a member of the International Association of Engineers IAENG and The Machine Intelligence Research Labs. Dr Maleh has made contributions in the fields of information security and privacy, Internet of things security, wireless and constrained networks security. His research interests include information security and privacy, Internet of things, networks security, information system, and IT governance. He has published over than 100 papers (book chapters, international journals, and conferences/workshops), 17 edited books, and 3 authored books. He is the editor-in-chief of the International Journal of Information Security and Privacy, and the International Journal of Smart Security Technologies (IJSST). He serves as an associate editor for IEEE Access (2019 Impact Factor 4.098), the International Journal of Digital Crime and Forensics (IJDCF), and the International Journal of Information Security and Privacy (IJISP). He is a series editor of Advances in Cybersecurity Management, by CRC Taylor \& Francis. He was also a guest editor of a special issue on Recent Advances on Cyber Security and Privacy for Cloud-of-Things of the International Journal of Digital Crime and Forensics (IJDCF), Volume 10, Issue 3, July–September 2019. He has served and continues to serve on executive and technical program committees and as a reviewer of numerous international conferences and journals such as Elsevier Ad Hoc Networks, IEEE Network Magazine, IEEE Sensor Journal, ICT Express, and Springer Cluster Computing. He was the Publicity chair of BCCA 2019 and the General Chair of the MLBDACP 19 symposium and ICI2C’21 Conference. He received Publons Top 1\% reviewer award for the years 2018 and 2019.
\end{IEEEbiography}

\EOD

\end{document}